\documentclass[lettersize,journal]{IEEEtran}
\usepackage{amsmath,amsfonts}
\usepackage{algorithmic}
\usepackage{algorithm}
\usepackage{array}
\usepackage[caption=false,font=normalsize,labelfont=sf,textfont=sf]{subfig}
\usepackage{textcomp}
\usepackage{stfloats}
\usepackage{url}
\usepackage{verbatim}
\usepackage{graphicx}
\usepackage{cite}
\usepackage{multirow}
\begin{document}

\title{A Compact Narrowband Antenna Design for RF Fingerprinting Applications}

\author{Sheng Huang,~\IEEEmembership{Member,~IEEE}, Cory Hilton,~\IEEEmembership{Graduate Student Member,~IEEE}, Steve Bush, Faiz Sherman,\\and Jeffrey A. Nanzer, \IEEEmembership{Senior Member,~IEEE}

\thanks{Manuscript received April 19, 2021; revised August 16, 2021.xxxxxxxxxxxxxxxxxxxxxxxxxxxxx
xxxxxx} 
		\thanks{This work was supported by the Procter \& Gamble Company. \textit{(Corresponding author: Jeffrey A. Nanzer)}}
\thanks{S. Huang, C. Hilton, and J. A. Nanzer are with the Department of Electrical and Computer Engineering, Michigan State University, East Lansing, MI 48824 USA (e-mail:
huang287@msu.edu; hiltonc2@msu.edu; nanzer@msu.edu)}
\thanks{S. Bush and F. Sherman are with the Procter \& Gamble Company, Cincinnati, OH 45202 USA.}
\thanks{xxxxxxxxxxxxxxxxxxxxxxxxxxxxxxxxxxxxxxxxxxxx}
\thanks{xxxxxxxxxxxxxxxxxxxxxxxxxxxxxxxxxxxxxxxxxxxx}
\thanks{xxxxxxxxxxxxxxxxxxxxxxxxxxxxxxxxxxxxxxxxxxxx}
\thanks{xxxxxxxxxxxxxxxxxxxxxxxxxxxxxxxxxxxxxxxxxxxx}
\thanks{xxxxxxxxxxxxxxxxxxxxxxxxxxxxxxxxxxxxxxxxxxxx}}
\markboth{Journal of \LaTeX\ Class Files,~Vol.~14, No.~8, August~2021}%
{Shell \MakeLowercase{\textit{et al.}}: A Sample Article Using IEEEtran.cls for IEEE Journals}

\IEEEpubid{}

\maketitle

\begin{abstract}
Radio frequency (RF) fingerprinting is widely used for supporting physical layer security in various wireless applications. In this paper, we present the design and implementation of a small antenna with low-cost fabrication that can be directly integrated with nonlinear passive devices, forming a passive RF tag providing unique nonlinear signatures for RF fingerprinting. We first propose a miniaturized meander line dipole, achieved by two folded arms on two sides of the substrate. This leads to antenna with a simple feeding structure and compact size, making it ideal for planar integration. Two antennas on Rogers 4350B and ultra-thin flexible Panasonic Felios are fabricated, achieving small size at 0.21$\times$0.06$\times$0.004\textit{\(\lambda_{0}\)$^3$} and 0.14$\times$0.1$\times$0.0008\textit{\(\lambda_{0}\)}$^3$ with realized gain of 1.87~dBi and 1.46~dBi. The passive tag consists of the proposed antenna structure and an integrated RF diode, and is further developed on both substrates, aiming to generate inter-modulation products (IMP) due to the nonlinearity of the diode, which can be used for device identification through classification algorithms. We investigate the nonlinearity of the designed tags for transmission at 15~dBm using two-tone signals. All tags produce a significant increased power at IMP frequencies at a range of 0.4~m. The tags on Rogers substrate provide around 23~dB IMP power increase and tags on flexible substrate embedded in lossy material provide around 16~dB power increase. These findings confirm that the proposed solution offers a simple passive tag design to support unique nonlinear signatures for RF fingerprinting applications in a simple, low-cost device. 
\end{abstract}

\begin{IEEEkeywords}
Radio frequency fingerprint, passive RF tags, nonlinearity, meander line antennas.
\end{IEEEkeywords}

\section{Introduction}
\IEEEPARstart{I}{n} modern wireless communication, ensuring robust physical layer security has become a fundamental necessity given the exponential growth of connected devices and the rapid expansion of the Internet of Things (IoT)~\cite{8744656,4343863,10008216}. Radio frequency (RF) fingerprinting is one of the most promising solutions for providing an extra security layer in wireless devices, as it utilizes the unique hardware characteristics to generate identifiable signatures. The RF fingerprinting technique enables device authentication, intrusion detection, and network security enhancement by distinguishing between legitimate and malicious transmitters \cite{8970312}.

Typically, RF fingerprinting relies on variations in hardware impairments, such as oscillator phase noise, device nonlinearity, and antenna mismatches, to produce distinct signatures. Most RF fingerprinting systems often require active signal processing and computationally intensive methods to extract and classify these signatures \cite{10130767,4211360}. However, such approaches can be inefficient for large-scale deployments, particularly in resource-constrained environments such as IoT networks. On the other hand, passive RF fingerprinting leverages inherent device-specific imperfections in the RF signals emitted by passive devices (e.g., RF identification (RFID) tags) for identification and authentication \cite{1549967,8736765}. Specifically, passive RF fingerprinting brings various key advantages such as: cost and energy efficiency supported by the passive tag without the need for an internal power source; unique hardware signatures generated by the manufacturing variances which are extremely difficult replicate; and ability of seamless integration with existing infrastructure, among others. To achieve scalability, research on deployment of compact, miniaturized antennas characterized by low-cost, ease of integration, and a simplified design for use in confined spaces is critical for practical applications.

Recently, various techniques have been proposed and studied in compact antenna designs, where one of the most popular methods involves the used of meander line element~\cite{5960759,6353124,7160701,5456169,10414395,5559341}. The meander line antenna is based on  periodic folded conductors, leading to a lowered resonance frequency with extended electrical length compared with the straight dipole antenna~\cite{99054,6171817}. Due to its easy design and efficient size reduction, many works on low-profile antenna designs have been reported in portable devices~\cite{1504825}, wearable antennas~\cite{9263312,9128053}, and RFID tags~\cite{6335459,5299014,6648416,9591355,7748548,7384427}. In~\cite{5299014}, a size-reduced passive RFID tag is proposed using a meander line antenna with periodically integrated LC elements for increasing read range. In~\cite{5979187}, a planar multi-band antenna is proposed for covering GSM, UMTS, and LTE with a compact size by introducing meander line monopole antennas. More recently, several designs on low-profile meander line based circular polarized antennas have been reported, but one of the draw backs is low gain due to low efficiency~\cite{9591355,6843861}, which is also a drawback in packaged small antennas~\cite{8758307}.

In this paper, we demonstrate a nonlinear low-profile and low-cost passive RF tag in the 2.4 GHz band based on a meander line structure that can retransmit increased in-band power at inter-modulation product (IMP) frequencies to support RF fingerprint identification. We first propose the design of a two-sided planar meander line which enables miniaturization and easy integration. Two designs are fabricated on conventional thin and ultra-thin flexible substrates that exhibit positive realized gain around 1.87~dBi and 1.46~dBi, respectively. Following that, we apply this design method to a passive tag and achieve a planar integration with a simple RF diode, which generates nonlinear signatures. The passive tags are also fabricated on both substrates and the designs on flexible substrate are further developed in device-embedded scenarios. Measurement results show that for low power two-tone transmitting signals, all tags produce significantly increased power at IMP frequencies. These findings confirm that the proposed passive antenna is a good candidate of low-end device for RF fingerprint technique.

This rest of this paper is organized as follows. In section II, we present the theoretical analysis and design process of the proposed compact planar meander line antenna, followed by simulated results and optimization on both substrates. Then we demonstrate the fabrication and measurement results. In section III, we show the passive tag design with integrated RF diode by applying the proposed method, including the developed planar design and the conformal design for embedded case. The final section demonstrates the tag fabrication and measurement results.

\section{The Design of Planar Meander Line Antenna}
\subsection{Designs on Conventional and Ultra-thin Flexible Substrate}
\begin{figure*}[t]
	\begin{center}
\subfloat[]{\includegraphics[width=3.4in]{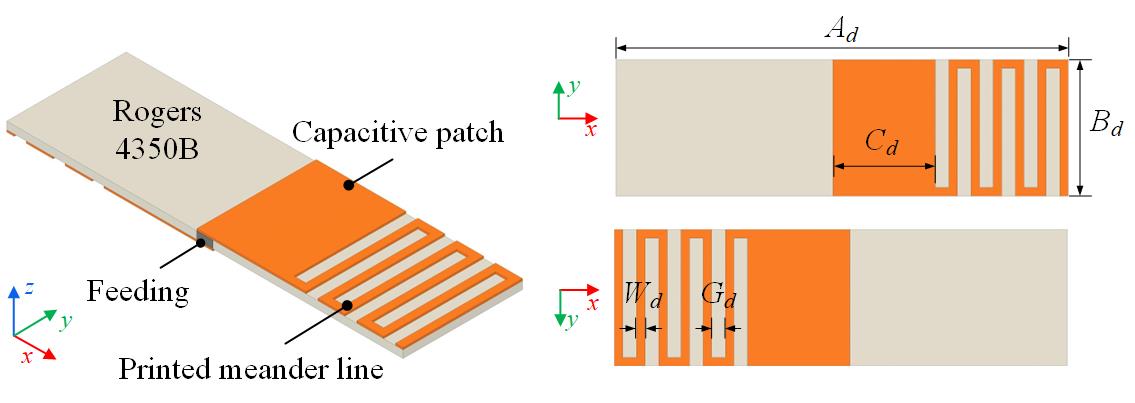}%
\label{figure1_first_case}}
\hfil
\subfloat[]{\includegraphics[width=3.4in]{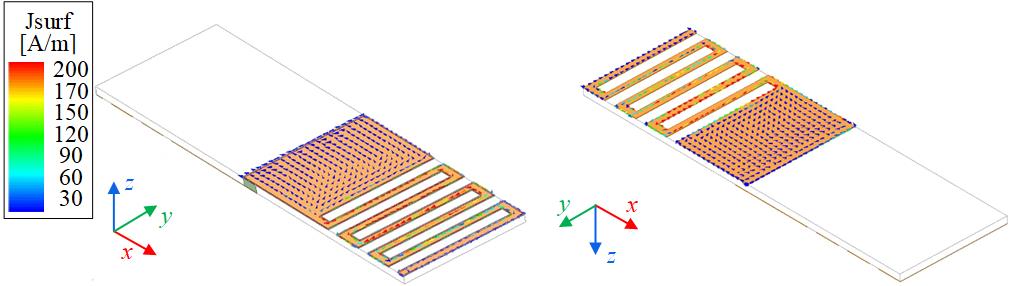}%
\label{figure1_second_case}}
\end{center}
\caption{The designed meander line dipole antenna width side feeding at the overlapped capacitive patch on Rogers 4350B with standard thin thickness at 0.508 mm. (a) The perspective, top, and bottom view and dimensions.  (b) The simulated current distribution on both printed metal at the resonance 2.4 GHz. The corresponding dimensions are \(A_{d}\) = 26.5, \(B_{d}\) = 8.2, \(C_{d}\) = 6, \(G_{d}\) = 0.8, \(W_{d}\) = 0.5 (unit:mm).}
\end{figure*}

\begin{figure}[t]
	\begin{center}
		\noindent
		\includegraphics[width=3in]{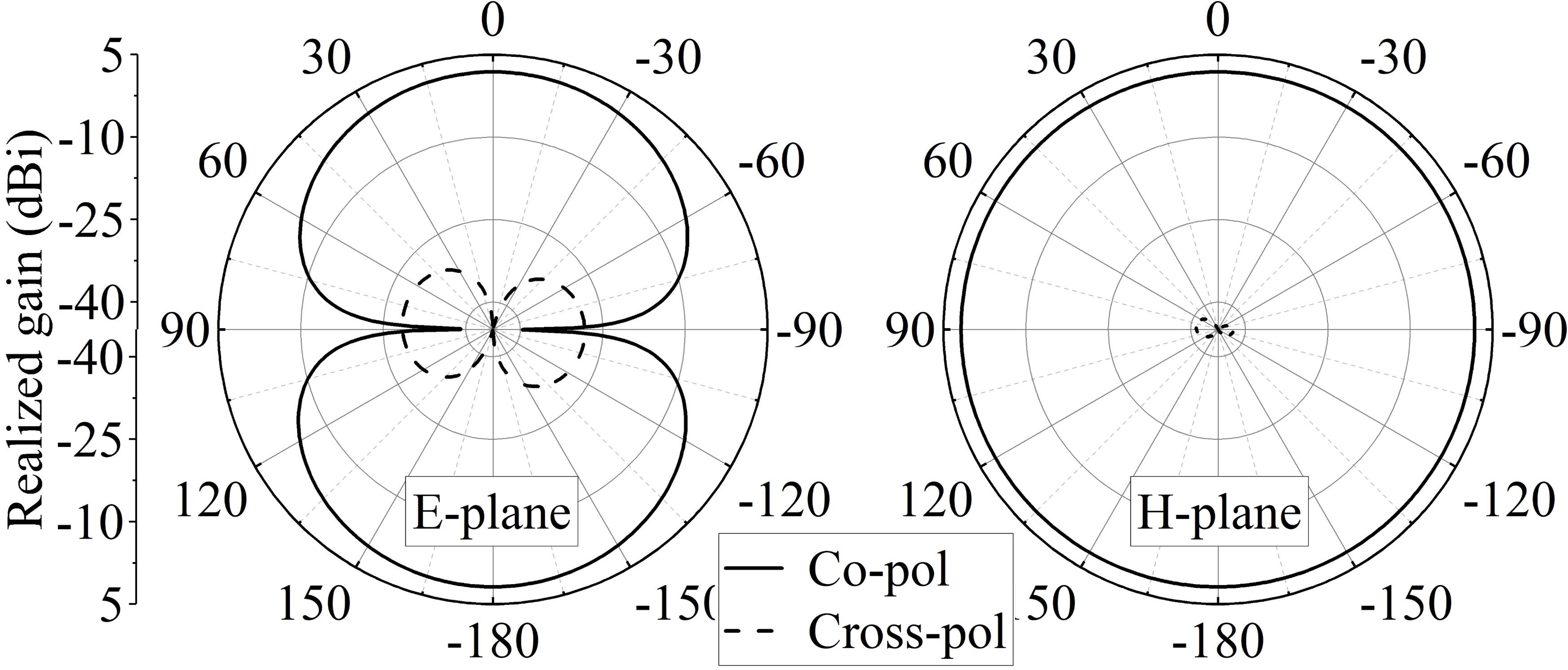}
		\caption{The simulated co-polarization and cross-polarization in E-and H-plane of the antenna in Fig. 1.}\label{Fig2}
	\end{center}
\end{figure}

\begin{figure}[t]
\centering
\subfloat[]{\includegraphics[width=2.9in]{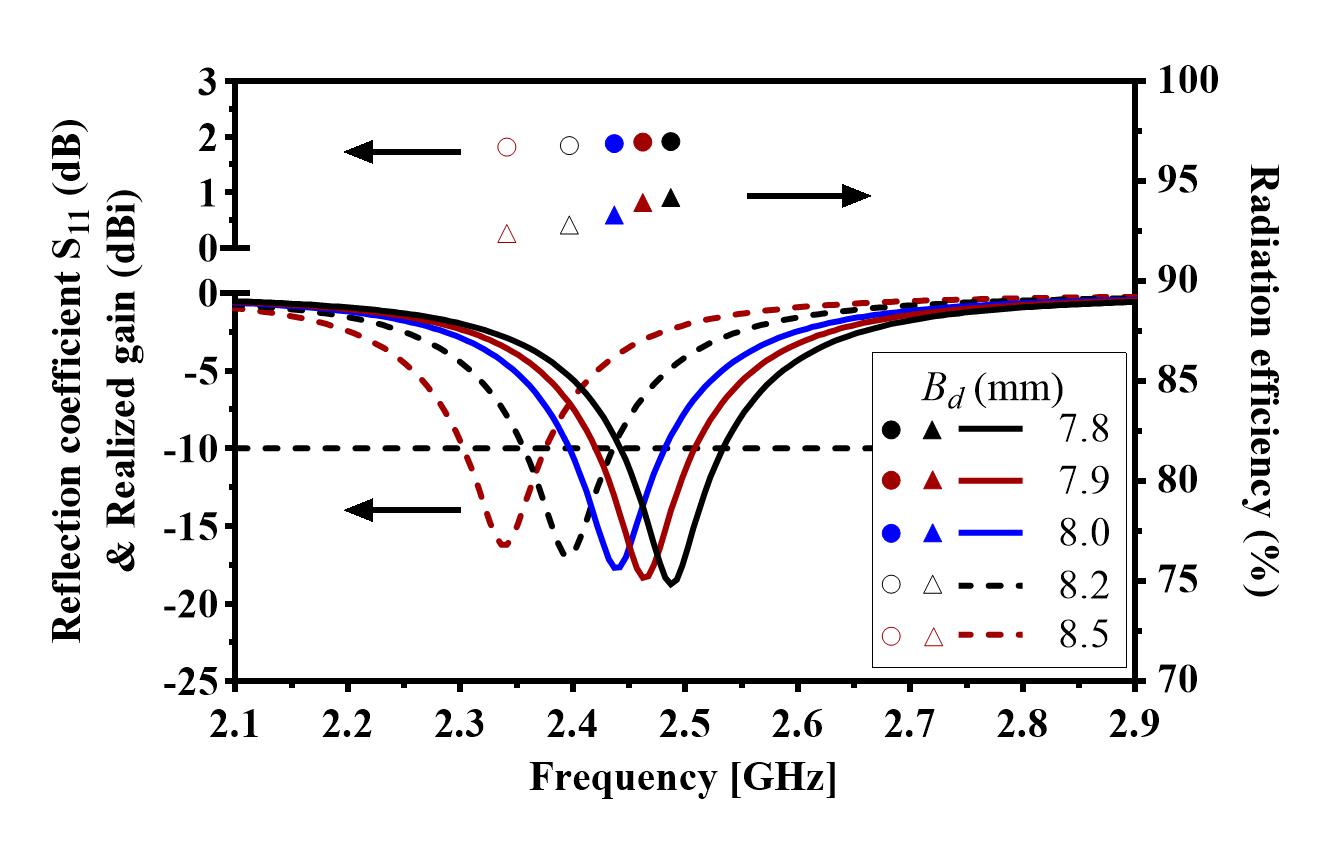}%
\label{figure9_first_case}}
\hfil
\subfloat[]{\includegraphics[width=2.9in]{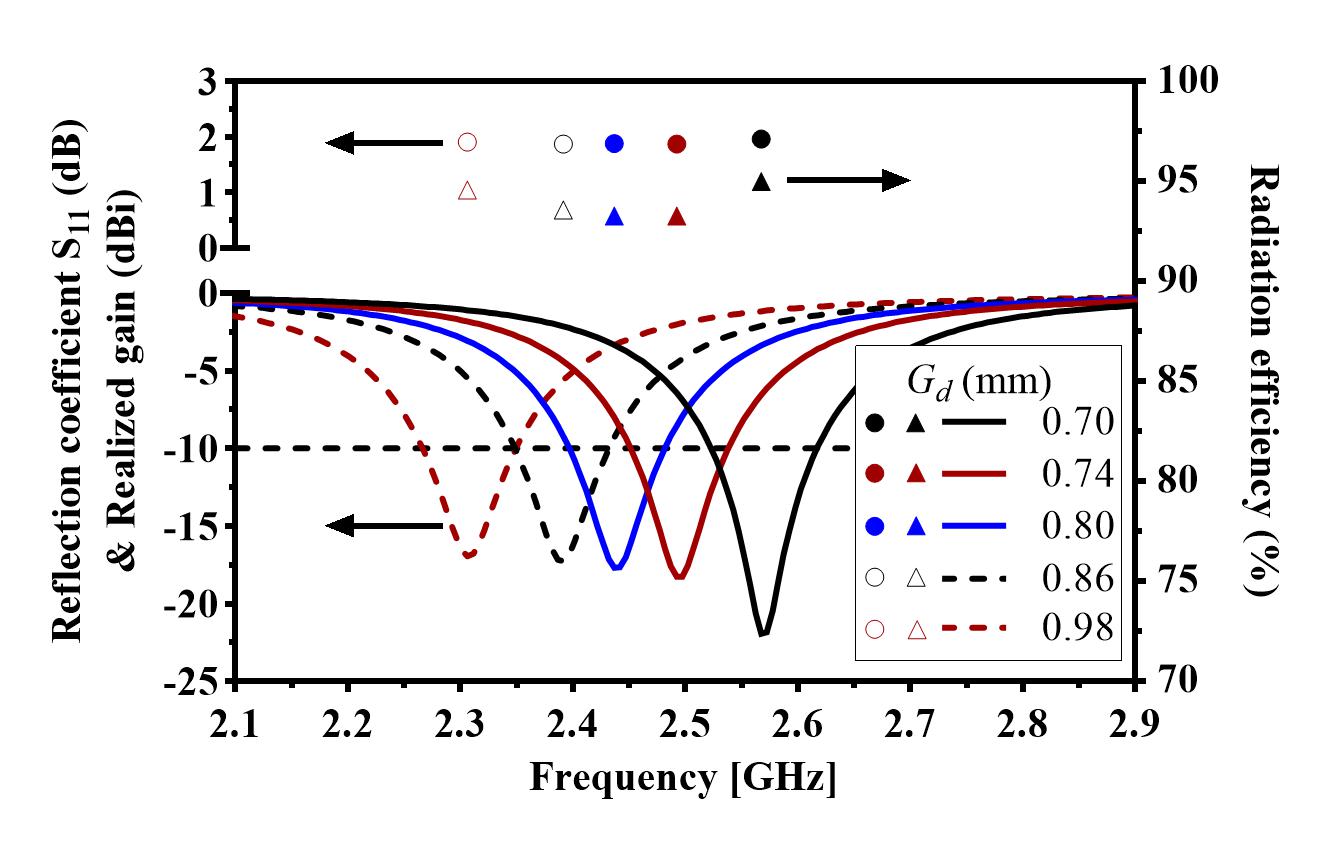}%
\label{figure9_second_case}}
\caption{The simulated reflection coefficient, realized gain, and radiation efficiency for (a) varied width \(B_{d}\), and (b) varied gap \(G_{d}\). (Other parameters are fixed as in table \ref{tab:table1})}
\label{figure9}
\end{figure}

The structure of the proposed planar meander line dipole antenna is shown in Fig. 1 (a). To achieve low-cost fabrication in a simple design process, the entire antenna is fabricated on a single layer of the Rogers substrate RO4350B with dielectric constant of 3.48 and dissipation factor of 0.0037. In our work, a standard thickness of 0.508 mm is used for a compact structure and light weight. We develop the meander line radiating elements on both top and bottom sides of the substrate, symmetrically distributed along the electrical polarization, leading to the overall planar size \(A_{d}\) $\times$ \(B_{d}\) around 0.21\textit{\(\lambda_{0}\)} $\times$ 0.06\textit{\(\lambda_{0}\)} (\textit{\(\lambda_{0}\)}=125 mm is the free space wavelength for 2.4 GHz). Each folded arm consists of three meander sections with length 2(\(B_{d}\)+\(W_{d}\)) around one quarter of the guided wavelength. A capacitive patch with length \(C_{d}\) inspired by \cite{4020418} is utilized on both sides, connecting the meander line section. It is worth noting that by doing this, an overlapped region which is created by the two capacitive strips in the middle of the substrate impacts not only the self-resonance of the meander line, but also helps with impedance matching. The overlapped capacitive strips also contribute to an easy way of antenna feeding and simple integration such as microstrip line and SMA. Other primary design parameters are the meander line width \(W_{d}\), gap \(G_{d}\) between the folded line, and the number \(N\) of turns for the resonance frequencies. 

Fig. 1 (b) plots the simulated surface current distribution on both sides of the printed metal by using ANSYS HFSS, showing that a typical meander line current distribution is achieved, where the current flow is in-phase on the horizontal segments in \(x\)-axis, and the current vanishes on the both end. The corresponding simulated realized gain radiation patterns in both E- and H-plane are plotted in Fig. 2, showing a realized gain of 1.85 dBi is obtained.
For the proposed meander line antenna, the self-resonance at 2.4 GHz is primarily influenced through the design of the full structure based principally on the width~\(B_{d}\) and the meander line gap~\(G_{d}\). This is because these two values mainly influence the electrical length and the total inductive reactance of the meander line. In Fig. 3, we explore the influence of the width~\(B_{d}\) and the meander line gap~\(G_{d}\) on impedance matching and radiation characteristics. Fig. 3 (a) shows the simulated reflection coefficient, realized gain, and radiation efficiency for width \(B_{d}\) from 7.8 mm to 8.5 mm. Stable fractional bandwidths of around 70 MHz can be maintained, and low conduction and dielectric loss result in high radiation efficiency above 90\%, leading to flat positive realized gain around 1.9~dBi. Fig. 3 (b) shows the simulated results for width \(G_{d}\) from 0.70 mm to 0.98 mm. Positive realized gain can also be maintained when the resonance is tuned from 2.3~GHz to 2.57~GHz.

\begin{figure}[t!]
	\begin{center}
		\noindent
		\includegraphics[width=3in]{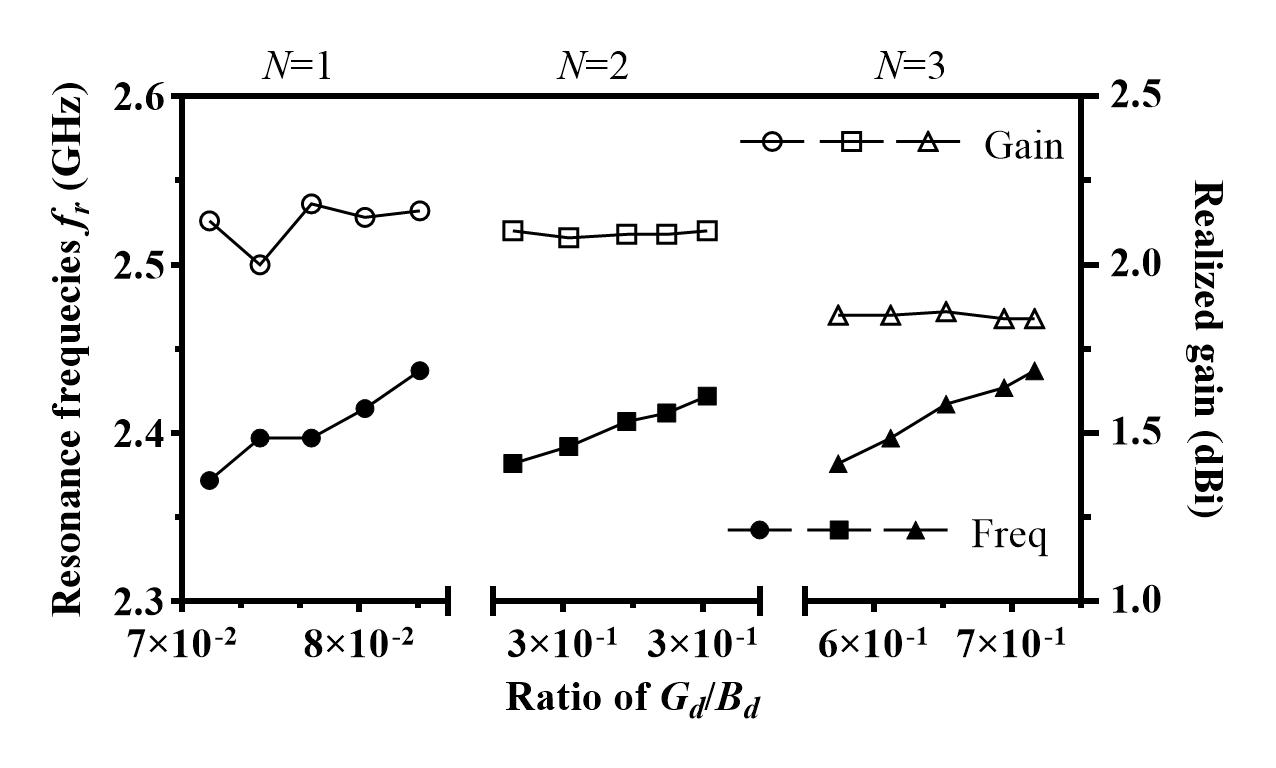}
		\caption{Investigation of the resonance and realized gain of the designed meander line antenna for the variation of the ratio of gap to line width and the number of meander segments.}\label{Fig1Label}
	\end{center}
\end{figure}
\begin{figure}[t!]
	\begin{center}
\subfloat[]{\includegraphics[width=3in]{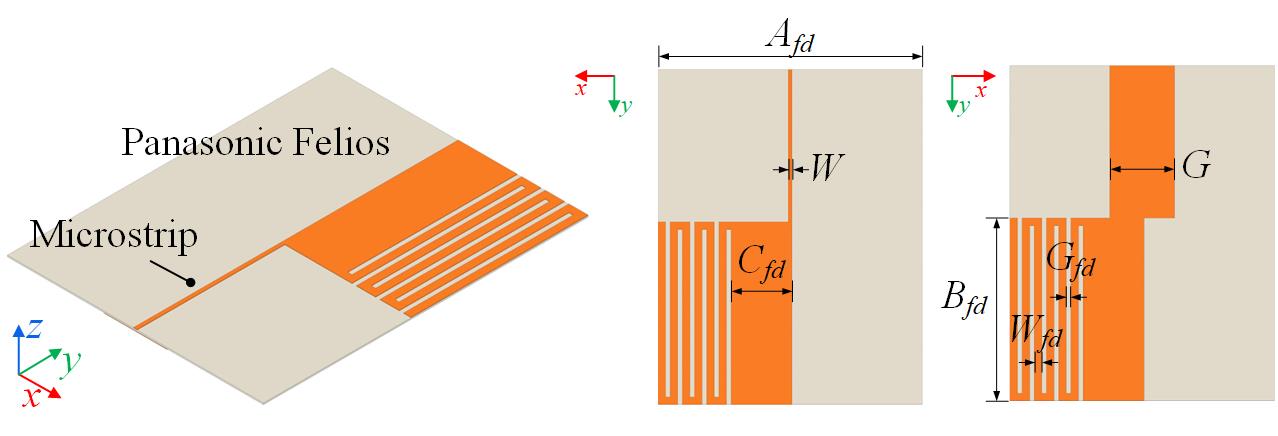}%
\label{figure5_first_case}}
\hfil
\subfloat[]{\includegraphics[width=2.4in]{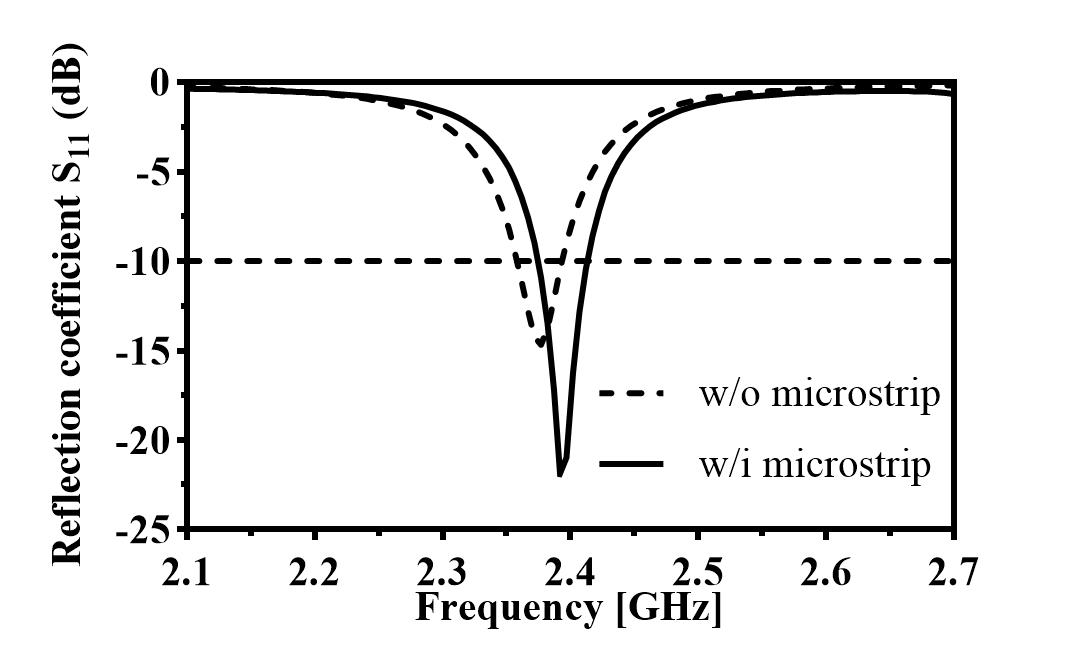}%
\label{figure5_second_case}}
\hfil
\subfloat[]{\includegraphics[width=3in]{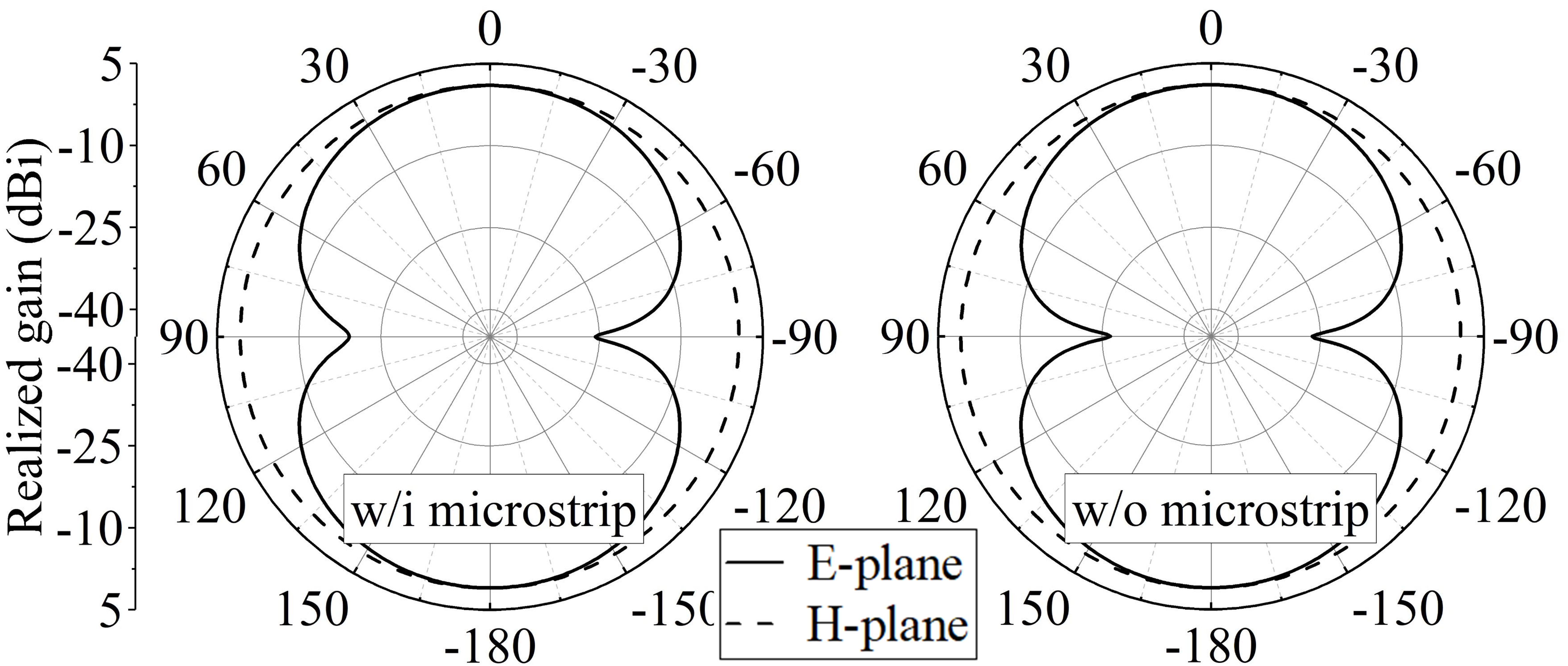}%
\label{figure5_second_case}}
	\end{center}\vspace{-1em}
\caption{The designed meander line dipole antenna with microstrip line feeding on ultra-thin flexible substrate with thickness at 0.1 mm at 2.4 GHz.  The corresponding dimensions are \(A_{fd}\) = 17.1, \(B_{fd}\) = 12, \(C_{fd}\) = 4, \(G_{fd}\) = 0.28, \(W_{fd}\) = 0.5, \(W\) = 0.246, \(G\) = 0.4. (unit:mm)}
\end{figure}

Fig. 4 demonstrates the relationship among the number of meander segments, resonance frequencies, and the ratio of line width \(W_{d}\) to the meander line gap \(G_{d}\). To demonstrate a few designs, we show the results for the resonance frequencies tuned around 2.4 GHz for a maximum of three meander segments ($1<N<3$), while other dimensions and properties including the substrate thickness and width \(W_{d}\) are fixed. Fig. 4 shows that for a given resonance frequency, increasing the number of segments is an effective way to reduce the antenna size and maintain the similar radiation performance. For example, to achieve the same resonance at 2.42 GHz, three meander segments design, having the ratio of \(G_{d}\)/\(W_{d}\) at 0.64 reduces the entire structure length \(A_{d}\) by about 32\%, compared to the one meander line segment design, having a ratio \(G_{d}\)/\(W_{d}\) of 0.078. It is also found that due to a smaller aperture size, a realized gain drop around 0.28 dBi is observed. Fig. 4 also shows that when increasing the number of meander sections, the meander line gap \(G_{d}\) is more sensitive to impedance matching. Therefore, a trade-off should be made among the size reduction and the number of meander line segments when considering the expected radiation performance and the fabrication tolerance. In our work, we choose three meander line segments in our design as it provides the required miniaturization for practical device sizes and the positive realized gain to support the retransmission of incident signals.

As shown in Fig. 5, we also developed the proposed antenna structure on low-cost flexible PCB Panasonic Felios R-775 series, which is widely used in commercial electronic device, automotive products and other IoT industrial application. The key features of this substrate include the stable dielectric properties and low loss (dielectric constant of 3.2 from 1~MHz to 10~GHz and dissipation factor at 0.002), excellent structural flexibility allows for bending and folding, and high thermal resistance, suitable for various environments. The meander line antenna is designed on an ultra-thin thickness of 0.1 mm. To show that the antenna can be simply fed, a microstrip line with 50~$\Omega$ characteristic impedance for width \(W\) of 0.246 mm is directly integrated onto the overlapped region. Fig. 5 (b) shows the simulated reflection coefficient of the designed structure with the optimized dimensions listed below for the resonance at 2.4 GHz. The simulated results for the meander line antenna  without the microstrip feeding is also plotted as a reference. It is seen that a slight resonance frequency shift around 20 MHz occurs which can be easily fixed by the physical meander line dimensions. The corresponding simulated realized gain radiation patterns are plotted in Fig. 5 (c); both antennas exhibit a maximum realized gain around 1.2 dBi.
\begin{figure}[t]
\centering
\subfloat[]{\includegraphics[width=1.3in]{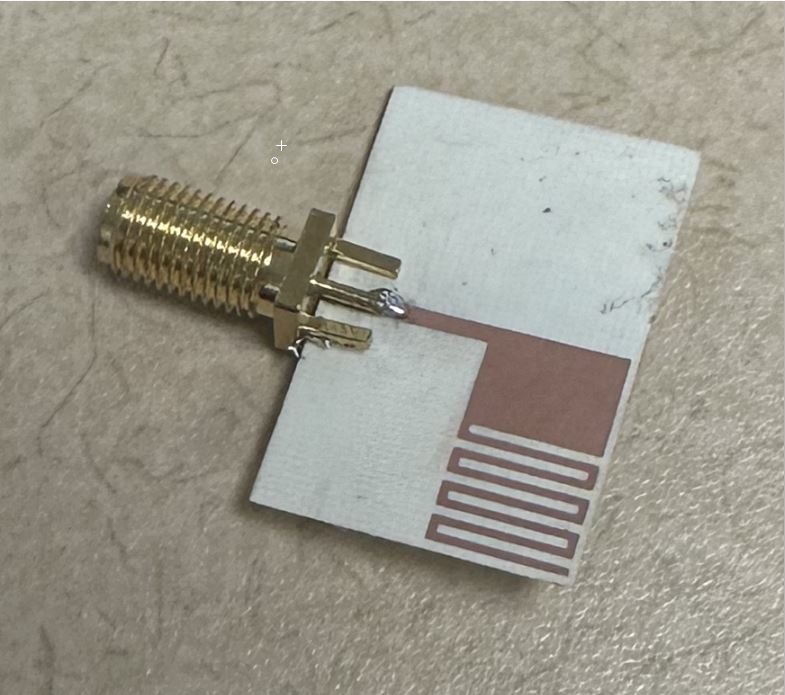}%
\label{figure9_first_case}}
\hfil
\subfloat[]{\includegraphics[width=1.48in]{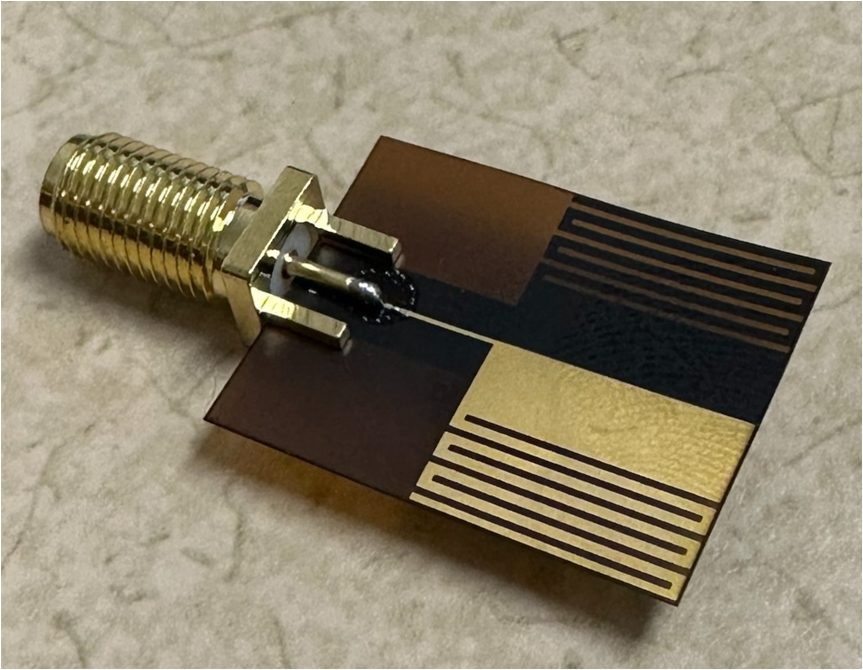}%
\label{figure9_second_case}}
\caption{The fabricated meander line dipole antennas on (a) Rogers 4350B, and (b) Panasonic Felios.}
\label{figure9}
\end{figure}
\begin{figure}[t]
\centering
\subfloat[]{\includegraphics[width=2.5in]{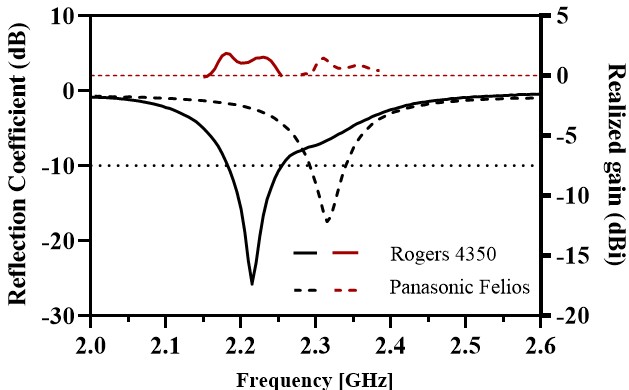}%
\label{figure9_first_case}}
\hfil
\subfloat[]{\includegraphics[width=3in]{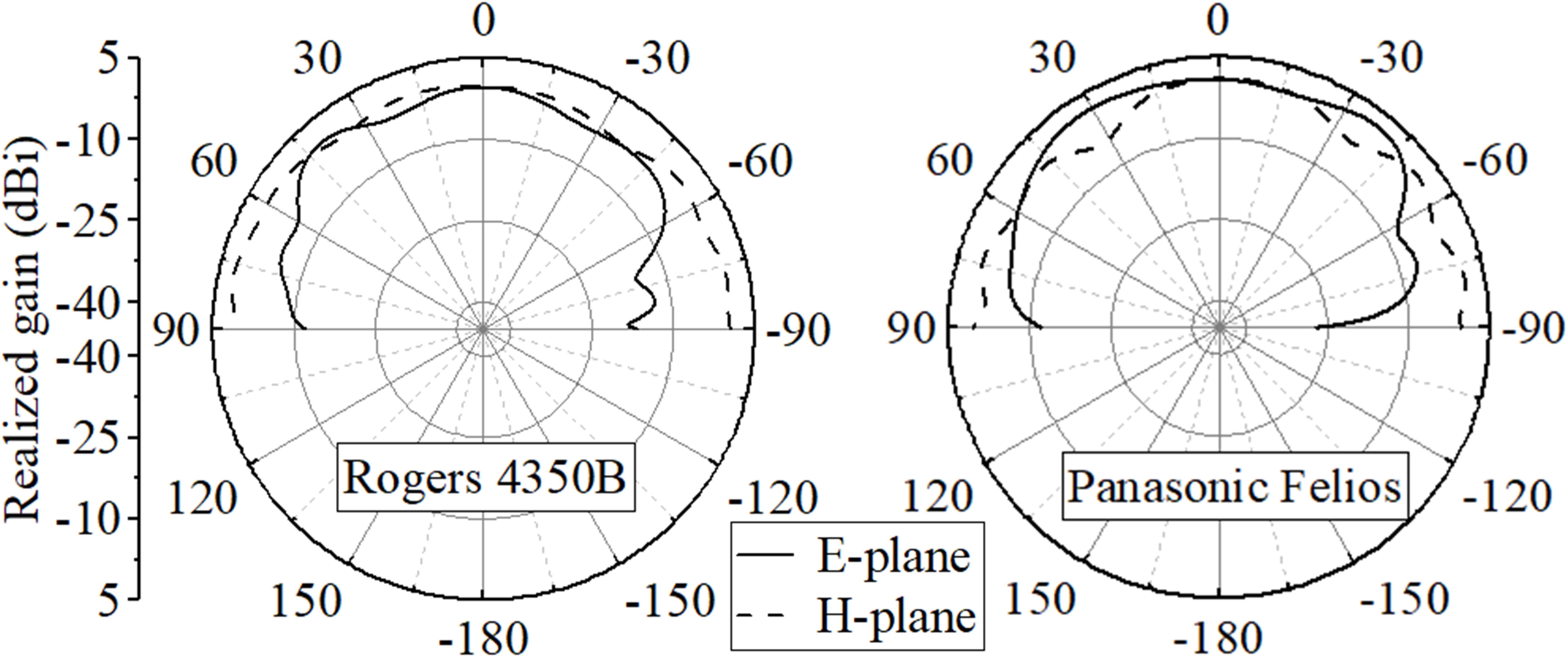}%
\label{figure9_second_case}}
\caption{The measurement results of the meander line antennas on Rogers 4350B and Panasonic Felios. (a) The measured reflection coefficient and realized gain. (b) The measured realized gain radaition patterns at the best resonance.}
\label{figure9}
\end{figure}

\subsection{Fabrication and Measurement Results}
The designed meander line dipoles were fabricated on both substrates, as shown in Fig. 6, to validate the radiation performance. The dimensions of the meander lines are optimized for 2.4 GHz self-resonance and antenna structures are fed by 50 ohm microstrip line with SMA feeding. The measured reflection coefficient and realized gain covering the fractional bandwidth are plotted in Fig. 7 (a). The resonance frequency of the two antenna are shifted to around 2.22 GHz and 2.32 GHz mainly due to interactions from the outer conductor of the closely placed SMA connector. The realized gain of the two antennas were measured with maximum value at 1.87 dBi and 1.46 dBi, respectively. The measured realized gain radiation patterns in both E- and H-plane are plotted in Fig. 7 (b), which closely match to the simulated patterns.

\section{The Design of Nonlinear Compact Passive Tag}
\subsection{Signature of the Proposed Tag}
In this section, we demonstrate the design of passive RF tags using the meander line antennas with integrated diodes to retransmit the signals with increased power at inter-modulation distortion (IMD) frequencies for RF fingerprinting. The nonlinearity of the RF diode mainly includes the carrier transport nonlinearity and the junction capacitance nonlinearity. The voltage-current (\(V\text{-}I\)) characteristics of the PN diode follows the Shockley equation
\begin{equation}
I = I_s \left( e^{\alpha{V}} - 1 \right)
\label{eq:diode_equation}
\end{equation}
where \(I_{s}\) is the reverse-saturation current, and $\alpha=q/nkT$, where $q$ is the elementary charge of an electron, $n$ is the ideality factor related to the diode structure, $k$ is Boltzmann's constant and $T$ is the absolute temperature in Kelvin \cite{pozar2021microwave}. For small signals, the current in the diode can be described using a Taylor series expansion as
\begin{equation}
    I \approx I_s \left( \frac{qV}{nkT} + \frac{1}{2!} \left( \frac{qV}{nkT} \right)^2 + \frac{1}{3!} \left( \frac{qV}{nkT} \right)^3 + \cdots \right)
    \label{eq:current_approximation}
\end{equation}
where the nonlearity coefficient of each term is determined by saturation current and ideality factor, and the second and third terms are the key factors that contribute to the nonlinearity characteristics. 

The junction capacitance \(C_{j}\) is nonlinear and can be described by 
\begin{equation}
    C_j = C_0 \left(1 + \frac{V}{V_b}\right)^{-m}
    \label{eq:junction capacitance}
\end{equation}
which can also be described using the Taylor expansion
\begin{equation}
  C_j = C_0 - mC_0 \frac{V}{V_b} + \frac{m(m+1)}{2} C_0 \frac{V^2}{V_b^2} + \ldots
\label{eq:junction capacitance taylor expan}
\end{equation}
where \(C_{0}\) is the junction capacitance at zero bias, \(V_{b}\) is the built-in voltage, and \(m\) is related to the diode type \cite{lin2011introduction}. The built-in voltage is affected by the doping concentration, and the component \(m\) is affected by the diode materials and design. Due to minor deviations in the manufacturing process, these characteristics uniquely determine the nonlinearity coefficients of each term, making the nonlinearity of every diode unique. 

Assuming that two input signals have the same power with equal voltage amplitude $V_{in}=V_{0}\cos(\omega_1 t)+V_{0}\cos(\omega_2t)$, the third order intermodulation distortion term, which manifests closest in frequency to the two input tones, can be given by
\begin{equation}
    IMD=\frac{3}{4} V_0^3 (\cos((2\omega_1 - \omega_2)t) + \cos((2\omega_2 - \omega_1)t))
\label{eq:third order imd}
\end{equation}
It is seen that increases in the power at the IMD frequencies are due to the nonliearity of the diode. Thus, a retransmitted third-order intermodulation signal will include minor nonlinear variations that are unique to each specific diode due to manufacturing differences. The nonlinear responses may then be used to uniquely identify the tags based on these small variations in power at the third order intermodulation frequency.

\begin{figure*}[t]
	\begin{center}
\subfloat[]{\includegraphics[width=1.5in]{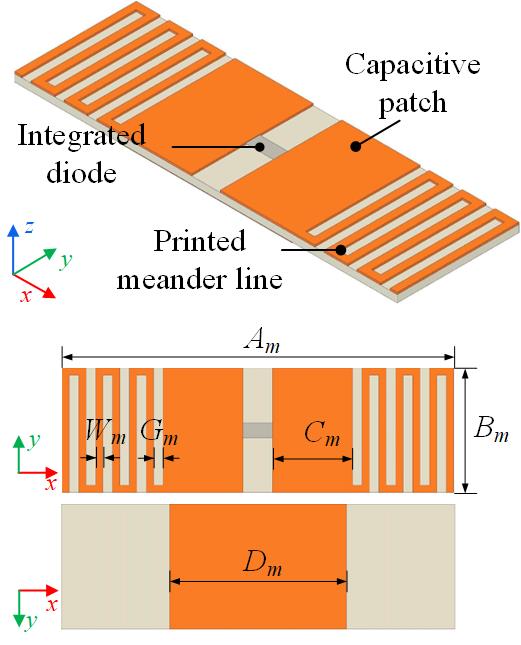}%
\label{figure9_first_case}}
\hfil
\subfloat[]{\includegraphics[width=1.7in]{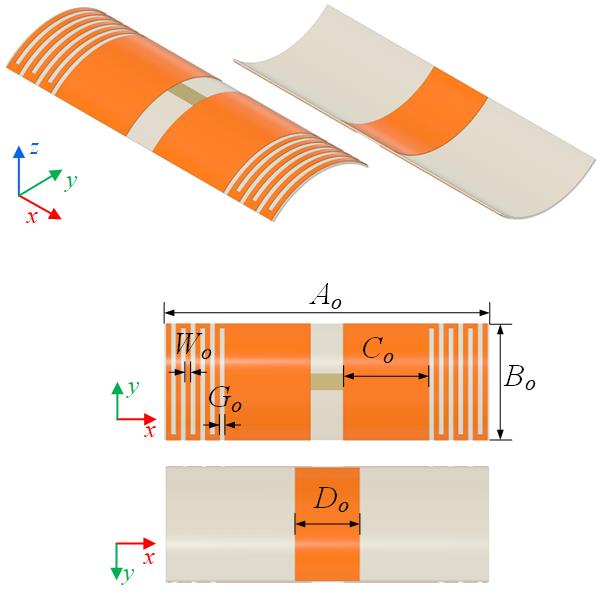}%
\label{figure9_second_case}}
\hfil
\subfloat[]{\includegraphics[width=1.7in]{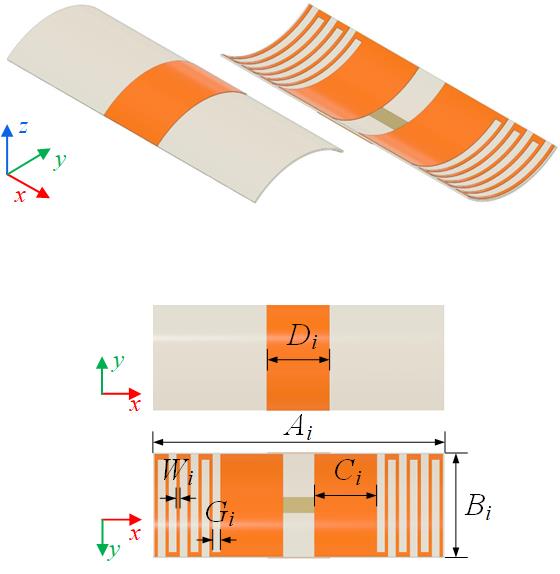}%
\label{figure9_second_case}}
\hfil
\subfloat[]{\includegraphics[width=1.5in]{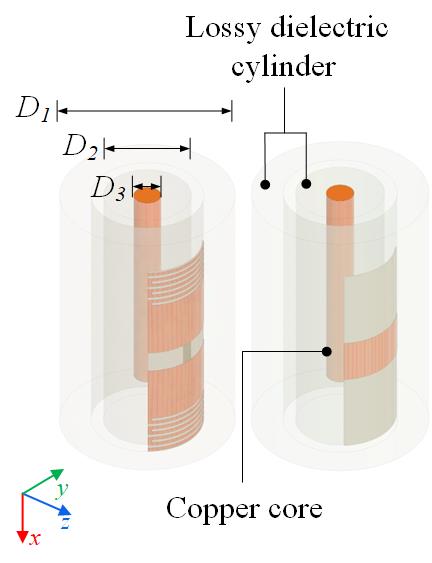}%
\label{figure9_second_case}}
	\end{center}\vspace{-1em}
\caption{The designed passive miniaturized meander line passive tags integrated with diode on planar substrate and cylindrical conformal structure on flexible substrate. (a) The design on Rogers RO4350B. (b) and (c) The designs on flexible substrate Panasonic Felios with curvature along \(y\)-axis. (d) The scenario of the embedded passive tag.}
\end{figure*}

\begin{figure*}[t]
	\begin{center}
\subfloat[]{\includegraphics[width=4.8in]{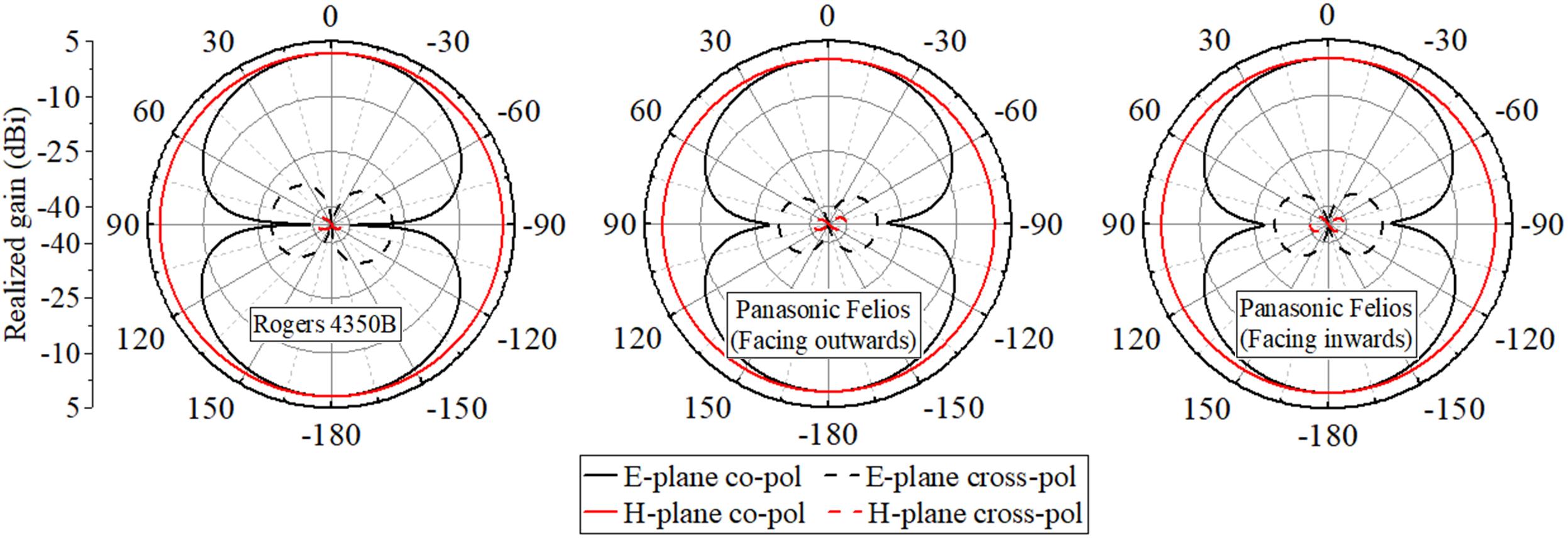}%
\label{figure9_first_case}}
\hfil
\subfloat[]{\includegraphics[width=2.32in]{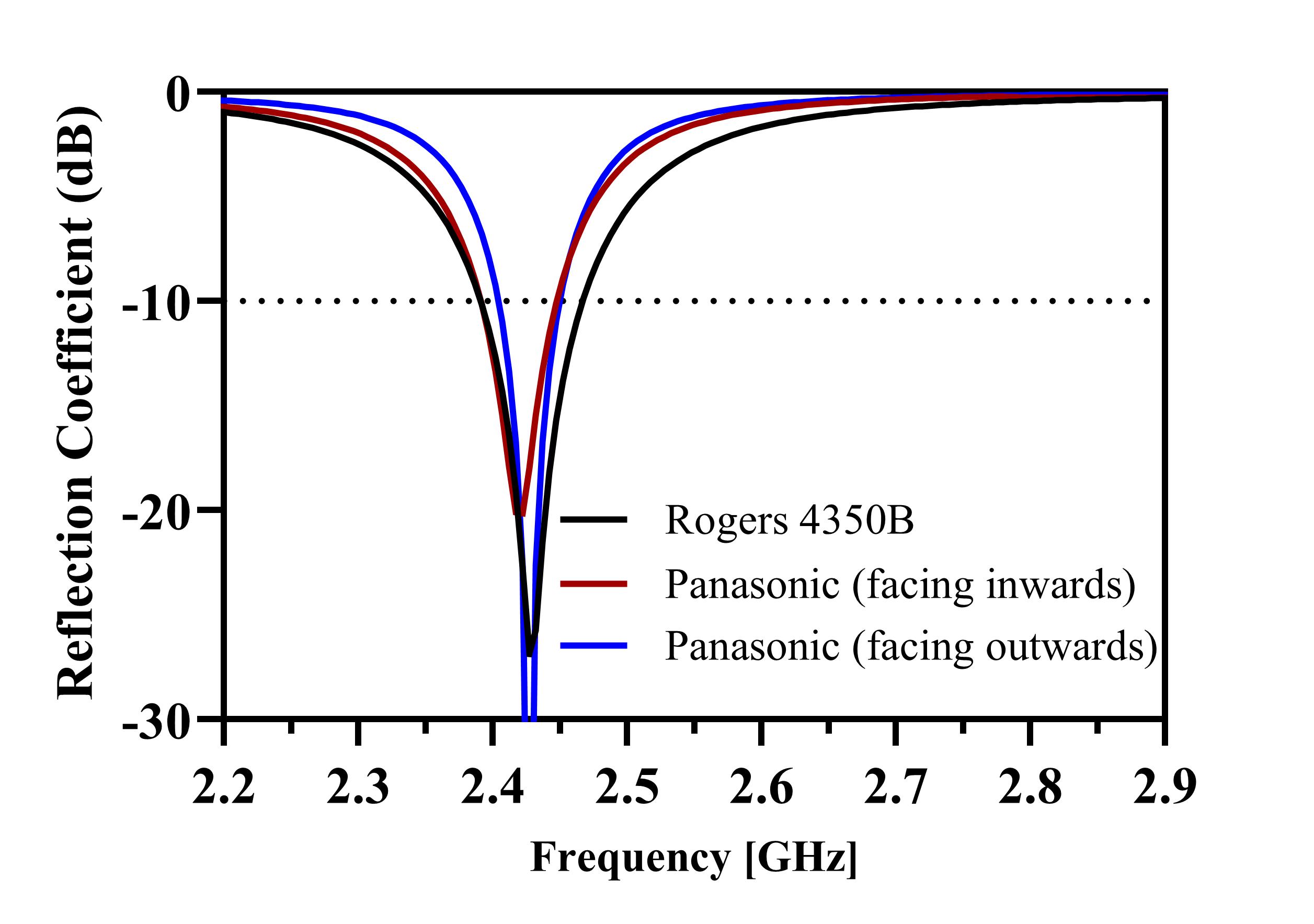}%
\label{figure9_second_case}}
	\end{center}\vspace{-1em}
\caption{The simulated matching and radtion performance the the designed passive tags at same resonance frequency 2.42 GHz. (a) The realized gain patterns of the design on RO4350B, on Panasonic Felio when embedded meander line facing outwards, and facing inwards. (b) The simulated resonance frequencies tunned at 2.42 GHz for three antennas. 
 (The passive antenna on Rogers4350B exhibits a maximum realized gain of 1.96 dBi, and the embedded antennas on Panasonic Felios exhibit realized gain of 0.7 dBi and 0.72 dBi, respectively.)}
\end{figure*}

\subsection{Design of the Proposed Tag}

As shown in Fig. 8 (a), we first implemented the meander line radiating elements to the same side of the substrate, and the diode is placed between the capacitive strips. Therefore, the planar configuration can be achieved on the thin substrate. On the back side, a ground patch is designed with length \(D_{m}\) to ensure the current flow. Fig. 8 (b) and (c) demonstrate the passive tag designs on Panasonic Felios flexible substrate in a conformal structure. This is then embedded in a lossy dielectric material as shown in Fig. 8 (d), which emphasizes that the proposed tag is suitable for complex and compact implementations. To demonstrate this, the passive tag is embedded in a two concentric cylinders with diameters of $D_1=16$ mm and $D_2=8$ mm, respectively, with identical thickness of 3 mm. The inner cylinder is filled with an copper core with a diameter \(D_{3}\) at 2 mm. The permittivity and the dissipation factor of the cylinders are 2.7 and 0.02, respectively, which are close to the value of typical commercially used plastic materials. Both tags on flexible substrate are place on the inner surface of the outer cylinder, which leads to a curved structure along the short side of the antenna with diameter of 10 mm. In this scenario, as the tags are surface mounted, we investigate two cases for the meander line elements when touching the dielectric surface, facing outwards ($+z$), and facing inwards ($-z$). This gives two tags conformal structures in Fig. 8 (b) and (c).

We study the impedance matching and the radiation characteristics of the three tags by using ANSYS HFSS. The initial designed dimensions follow the previous section. As the integrated diode supports the current flow between two arms, we set the lumped-port excitation to feed the current. Three antennas are tuned for the same resonance at 2.42 GHz which is mainly achieved by reducing the gap between the meander lines and the width of the entire antenna. The optimized dimensions are listed in Table \ref{tab:table1}. The simulated reflection coefficients are plotted in Fig. 9 (b).  In terms of the tags using flexible substrate, the embedded material leads to a more compact structure as it reduces the effective wavelength along the meander line. The simulated realized gain of three antenna are shown in Fig. 9 (a). The embedded material results in an expected gain drop of around 0.7 dBi, contributing about 0.75 dBi loss comparing with the measurement of dipole on the same substrate.
\begin{table}[t]
\caption{Dimensions of the Investigated Antennas for the Resonance at 2.42 GHz   (Unit: mm)\label{tab:table1}}
\centering
\renewcommand{\arraystretch}{1.2}
{
\begin{tabular}{cccccc} \hline \hline
\(A_{m}\)&\(B_{m}\) &\(C_{m}\) & \(D_{m}\) &\(G_{m}\) &\(W_{m}\)\\  
28& 8.2& 6& 12& 0.8& 0.5\\\hline
\(A_{mo}\)&\(B_{mo}\) &\(C_{mo}\) & \(D_{mo}\) &\(G_{mo}\) &\(W_{mo}\)\\
20& 7.76& 5.3& 4& 0.3& 0.3\\\hline 
\(A_{mi}\)&\(B_{mi}\) &\(C_{mi}\) & \(D_{mi}\) &\(G_{mi}\) &\(W_{mi}\)\\
18.6& 7.41& 4& 4& 0.2 & 0.5\\\hline\hline
\end{tabular}
}
\end{table}

\subsection{RF Tag Fabrication and Measurement}

\begin{figure}[t]
\centering
\includegraphics[width=3.4in]{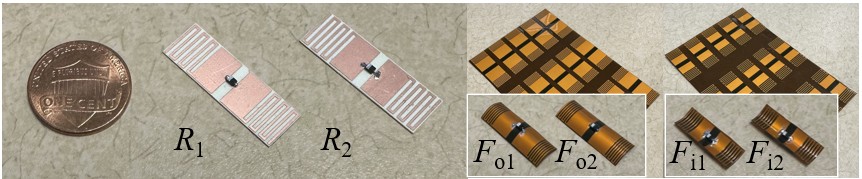}%
\hfil
\caption{The fabricated passive tags in Wifi 2.4 GHz frequency band with integrated diode on Rogers 4350B (\(R_{1}\) and \(R_{2}\)) and ultra-thin Panasonic Felios \((F_{o1}\), \(F_{o2}\), \(F_{i1}\), and \(F_{i2}\)).}
\label{figure9}
\end{figure}

\begin{figure*}[b]
\centering
\subfloat[]{\includegraphics[width=3in]{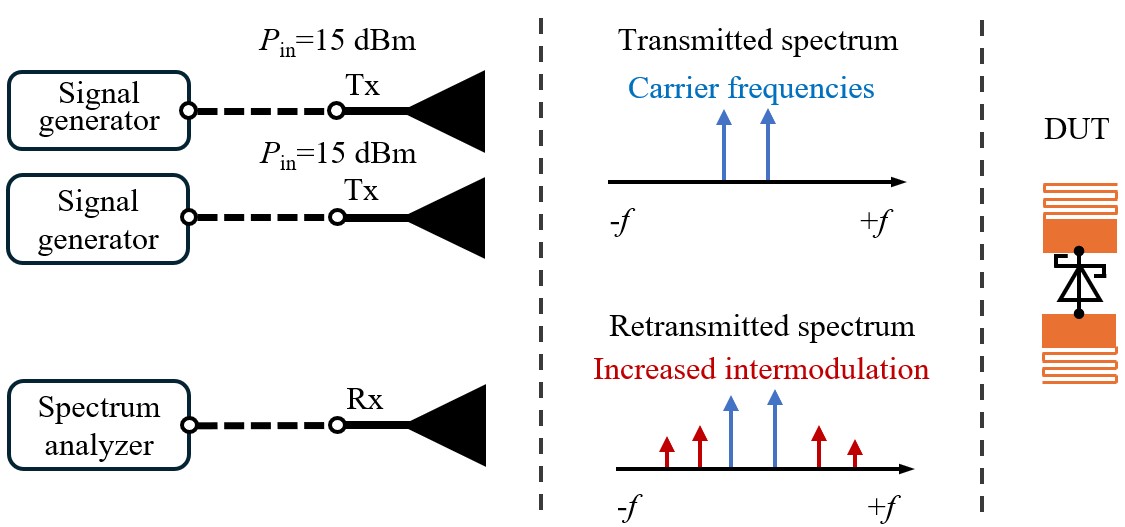}}%
\label{figure8_first_case}
\hfil
\subfloat[]{\includegraphics[width=3in]{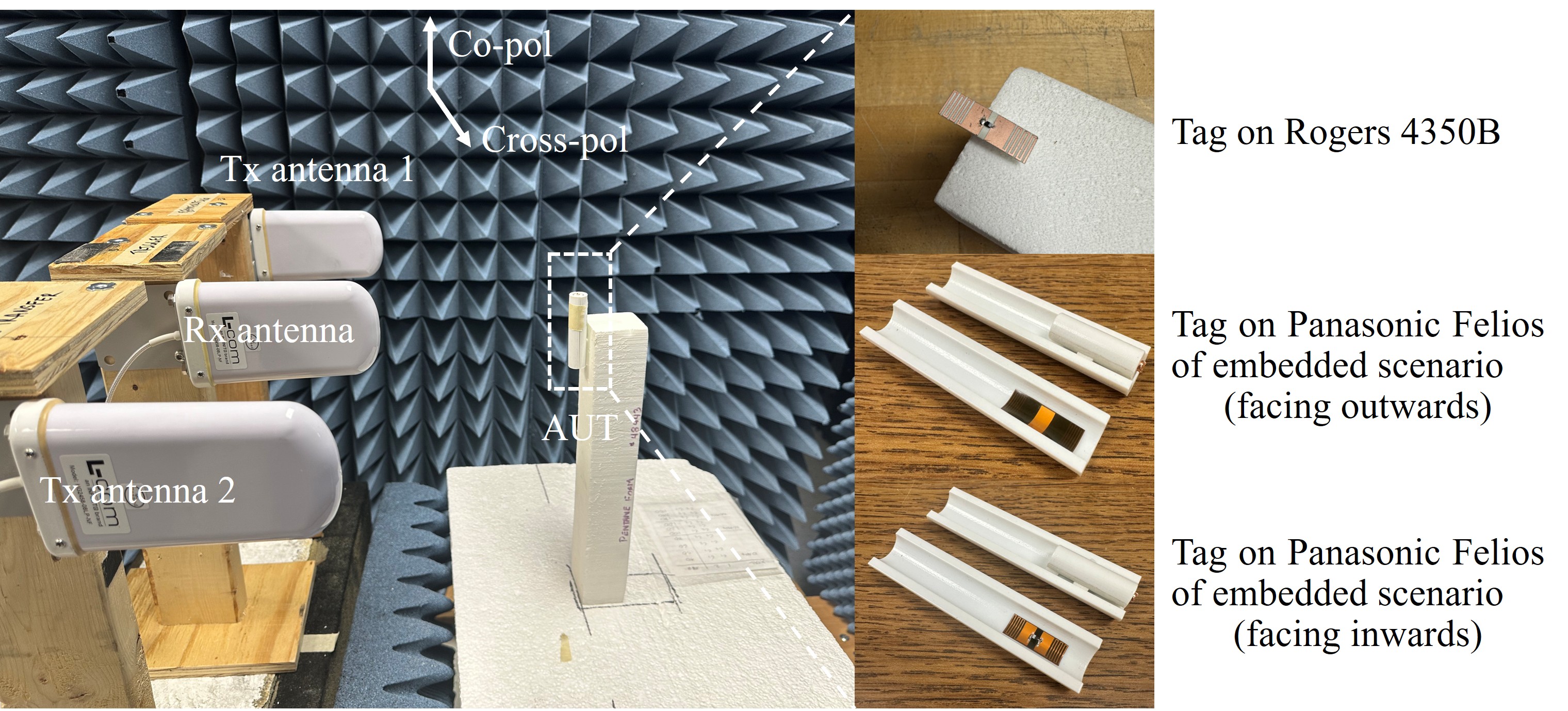}%
\label{figure9_second_case}}
\hfil
\caption{The measurement for the IMP. (a) The diagram of the measurement. (b) The measurement setup of Tx and Rx antennas and DUT. (The commercial wideband log periodic antennas with nominal 8 dBi realized gain from 2.4 GHz to 5.6 GHz are used at both transmitter and receiver.)}
\label{figure8}
\end{figure*}
To validate our hypothesis that the proposed passive tag is able to generate and retransmit the increased intermodulation distortion at harmonics, we fabricated the antennas on both the Rogers substrate and the flexible PCB, as shown in Fig.~\ref{figure9}. Two tags \(R_{1}\) and \(R_{2}\) fabricated on RO4350B using chemical etching are shown in Fig. 10 (a). The diode was soldered directly on both capacitive strips. Fig. 10 (b) shows the fabricated tags on flexible PCB, where the tags \(F_{o1}\) and \(F_{o2}\) were designed for the radiating element facing outward, and \(F_{i1}\) and \(F_{i2}\) were designed for facing inward. The final curved structure was achieved by fixing the tags onto a 1 cm diameter cylinder.
Fig. 11 shows the measurement setup to observe the IMP, where Fig. 11 (a) is the diagram of the measurement. For the two transmitting antennas, commercial wideband log periodic antennas (L-Com HG2458-08LP-NF) were used to transmit monotone continuous wave signals from two signal generators (Agilent N5183A and Keysight E8267D). Separate signal generators were used to ensure that any nonlinearities observed in the received spectrum were due to the tag and not the transmitting signal. The input RF power \(P_{in}\) of the transmitting antennas was identical at 15 dBm, with a 10 KHz frequency separation. The same type of log periodic antenna was positioned between the transmitting antennas to receive the retransmitted power from the device under test and Keysight microwave analyzer N9952A was used for signal analyzing. The passive tags were measured at both co-polarization and cross-polarization orientations. Two control measurement were also conducted, including a piece of copper with identical size of with the tags, and a measurement with only the standing foam structure used to hold the tags.

The measured received frequency spectra over 80 KHz frequency span are plotted in Fig. 12, where Fig. 12 (a) is the spectrum for tag \(R_{1}\) and Fig. 12 (b) is the spectrum for tag \(R_{2}\). The tag \(R_{1}\), which had best resonance around 2.41 GHz, was firstly measured. The measured received frequency spectrum for the copper (grey) and standing foam (blue) are also plotted. All measured power levels are normalized by setting the peak power of the carrier frequencies at 0 dBm. We can see that for the measured frequency spectrum of copper and standing foam, only the carrier frequencies can be observed, as expected since there are no nonlinear devices in the system. This result is also reflected in the cross-polarization of the tags. On the other hand, the increased power level of IMP frequencies is obtained for the co-polarization of both tags, reaching at -55.9 dB (\(R_{1}\)) and -57.8 dB (\(R_{2}\)), yielding around 23 dB over the noise floor.

\begin{figure*}[t]
\centering
\subfloat[]{\includegraphics[width=2.2in]{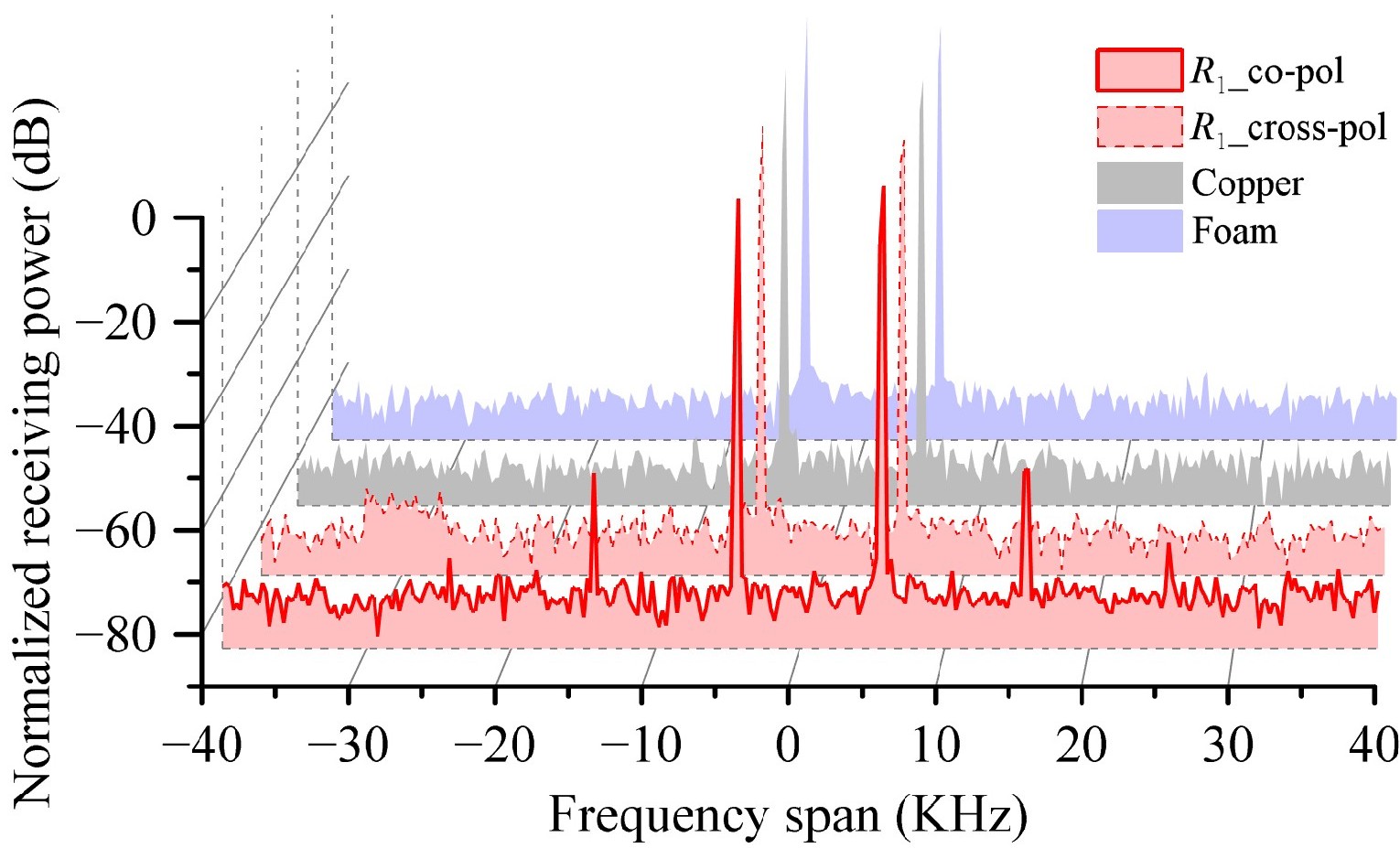}%
\label{figure12_first_case}}
\hfil
\subfloat[]{\includegraphics[width=2.2in]{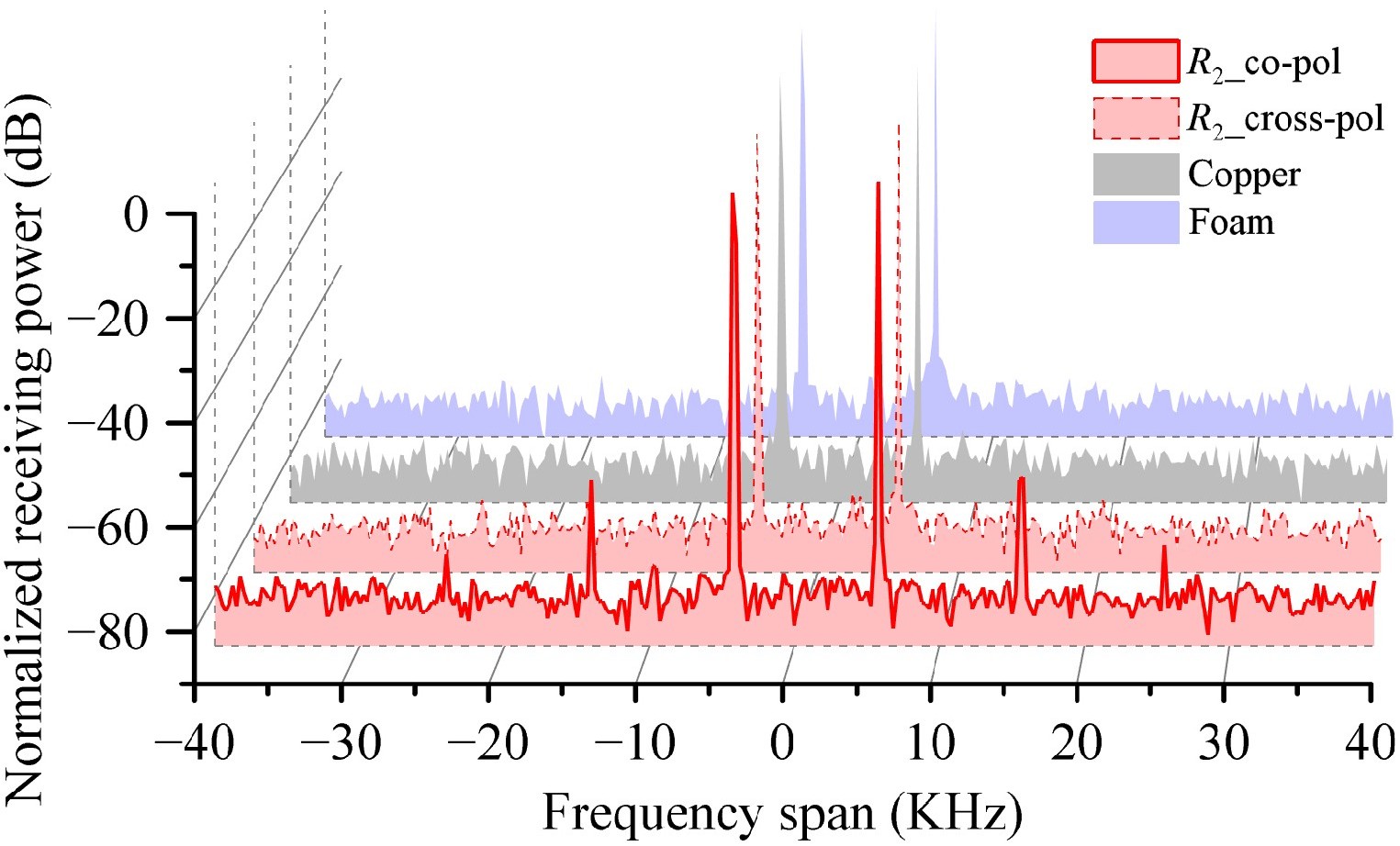}%
\label{figure12_second_case}}
\hfil
\subfloat[]{\includegraphics[width=2.2in]{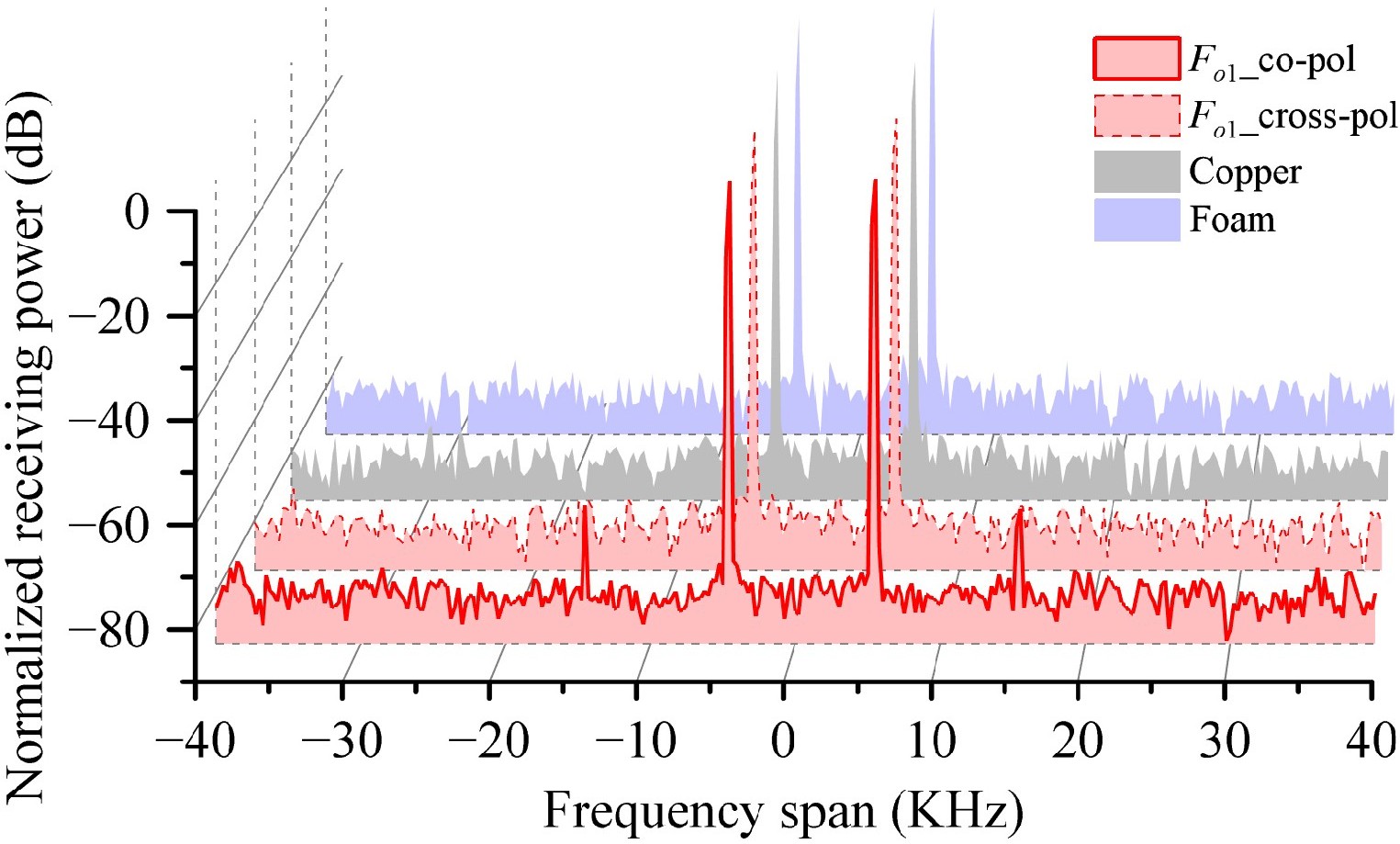}%
\label{figure12_third_case}}
\hfil
\subfloat[]{\includegraphics[width=2.2in]{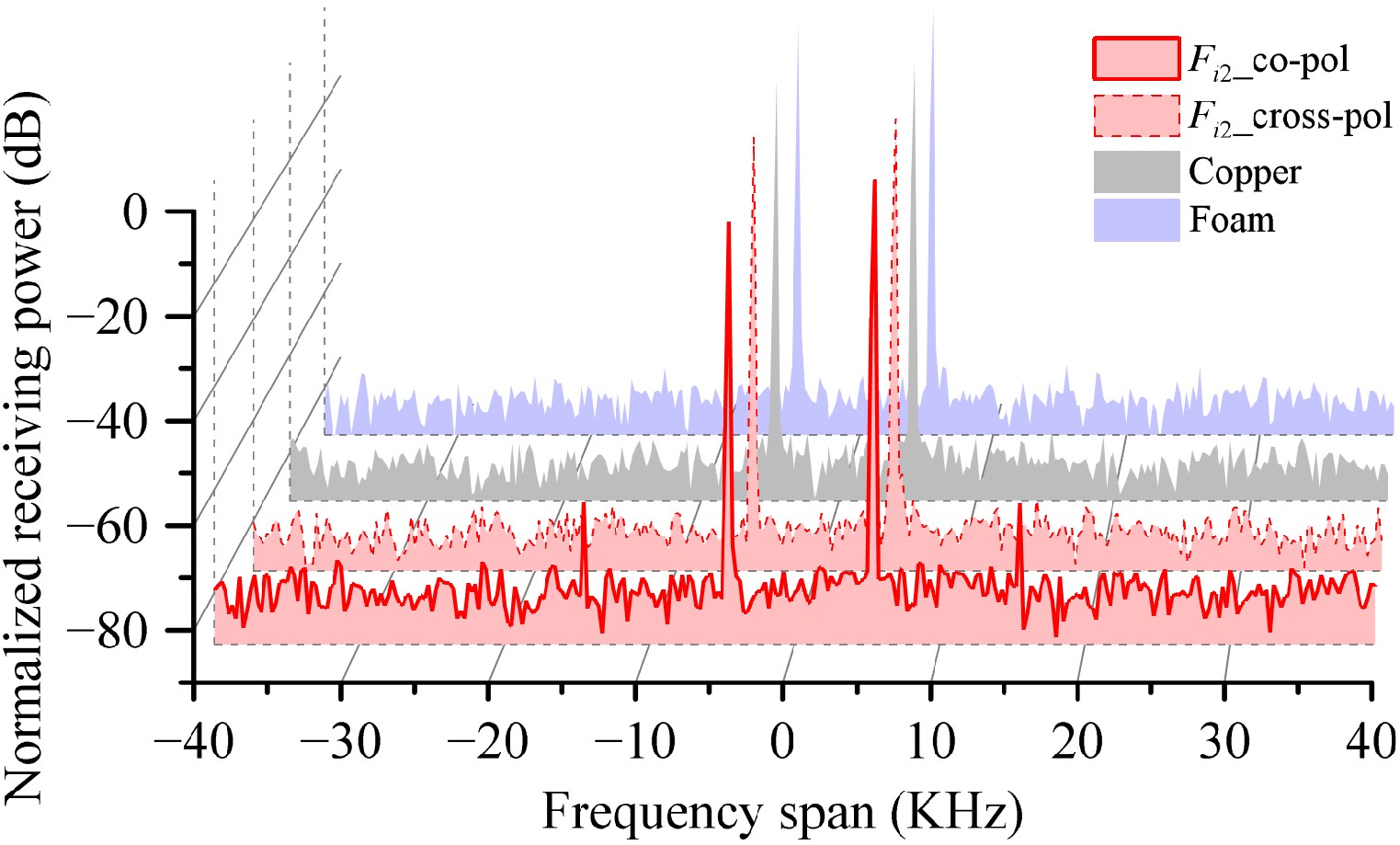}%
\label{figure12_fourth_case}}
\hfil
\subfloat[]{\includegraphics[width=2.2in]{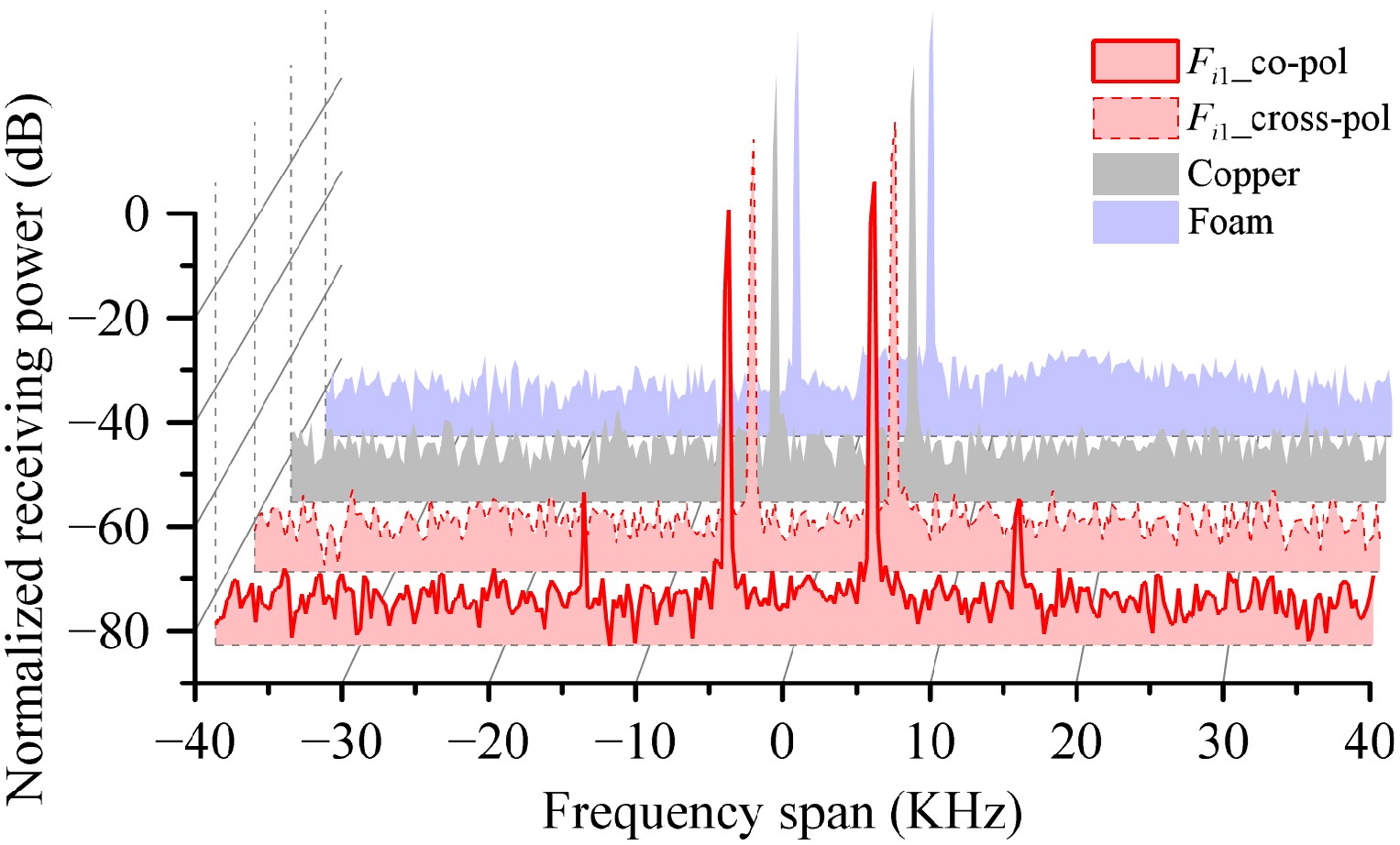}%
\label{figure12_fifth_case}}
\hfil
\subfloat[]{\includegraphics[width=2.2in]{Tag_Fi2.jpg}%
\label{figure12_sixth_case}}
\hfil
\caption{The measured frequency spectrum of fabricated passive tags on Rogers 4350B and ultra-thin Panasonic Felios in 80 KHz frequency span. (a) \(R_{1}\), (b) \(R_{2}\), (c) \(F_{o1}\), (d) \(F_{o2}\), (e) \(F_{i1}\), (f) \(F_{i2}\). (The corresponding designed resonance frequencies are 2.41 GHz, 2.42 GHz, 2.405 GHz, 2.415 GHz, 2.425 GHz, and 2.475 GHz. The measured results show that increased power at IMP frequencies can only be observed when the passives tags are in co-polarization with transmitting antennas, carrying unique signatures due to low-cost nonlinear diode.)}
\label{figure12}
\end{figure*}

Fig. 12 (c) and (d) demonstrate the measured received frequency spectrum for the tags \(F_{o1}\) and \(F_{o2}\) on ultra-thin flexible PCB of the embedded scenarios when the radiating meander line is facing outwards. The best resonances of \(F_{o1}\) and \(F_{o2}\) are 2.405 GHz and 2.415 GHz, respectively. Similar trends are observed, namely that no increased power at IMP is observed for the cases of cross-polarization, copper, and standing form. The received frequency spectrum for co-polarization of the tags shows obvious first IMP increase at -63.2 dB (\(F_{o1}\)) and -63.8 dB (\(F_{o2}\)). These values are about 7 dB lower than the \(R_{1}\) and \(R_{2}\) due to the dielectric loss of the bulky packaging material. Fig. 12 (e) and (f) show the measured received frequency spectrum for the tags \(F_{i1}\) and \(F_{i2}\) on ultra-thin flexible PCB of the embedded scenarios when the radiating meander line is facing inwards. The increased power at first IMP of \(F_{i1}\) and \(F_{i2}\) were measured at -60.3 dB and -62.6 dB, respectively. Based on the measurement results, it can be concluded that the proposed small passive tags on both the Rogers substrate and ultra-thin PCB are able to retransmit the increased IMP only due to the non-linearity of the diode, which enables the tags to carry a unique signature. It also shows that the observation of the IMP can be achieved with low transmitting power and low gain antennas, suitable for low power and low-cost applications. The measurement details are summarized in Table \ref{tab:table2}.

\begin{table}[t!]
\caption{Measured and Simulated Results of Three Antennas
\label{tab:table2}}
\centering
\renewcommand{\arraystretch}{1.2} 
\resizebox{\linewidth}{!}
{
\begin{tabular}{cccccc} \hline \hline

\multicolumn{1}{c}{\begin{tabular}[c]{@{}c@{}}Passive\\tag\end{tabular}}& \multicolumn{1}{c}{\begin{tabular}[c]{@{}c@{}}\(f_{r}\)\\(GHz)\end{tabular}}&
\multicolumn{1}{c}{\begin{tabular}[c]{@{}c@{}}\(f_{c1}\)\\ (GHz)\end{tabular}} &
\multicolumn{1}{c}{\begin{tabular}[c]{@{}c@{}}\(f_{c2}\)\\ (GHz)\end{tabular}} &
\multicolumn{1}{c}{\begin{tabular}[c]{@{}c@{}}IMP\\ (dB)\end{tabular}}&
\multicolumn{1}{c}{\begin{tabular}[c]{@{}c@{}}SNR\\ (dB)\end{tabular}} \\ \hline

\(R_{1}\)& 2.41& 2.409995& 2.410005& -55.9&23.69\\  

\(R_{2}\)& 2.42& 2.419995& 2.420005& -57.8&23.09\\ 
\(F_{o1}\)& 2.405& 2.405995& 2.405005& -63.2&17.34\\ 
\(F_{o2}\)& 2.415& 2.414995& 2.415005& -63.8&16.45\\
\(F_{i1}\)& 2.425& 2.424995& 2.425005& -60.3&16.23\\
\(F_{i2}\)& 2.475& 2.474995& 2.475005& -62.6&14.98\\\hline\hline
\end{tabular}
}
\begin{flushleft}
*\(f_{c1}\) and \(f_{c2}\) are the carrier frequencies of two tone transmitting signals. The SNR is calculated by comparing the measured power at IMP frequencies retransmitted by the designed tag with the measured power of standing foam.
\end{flushleft}
\end{table}
\vspace{-0.001em} 

\section{Conclusion}
In this paper, we introduced a compact meander line antenna design that can be easily integrated in planar circuits, achieving both miniaturization and superior radiation performance. The proposed method was applied to passive tag design integrated with and RF diode, offering a simple and cost-effective solution for RF fingerprinting of low-end devices in low power wireless applications. The designed tags leverage the nonlinearity of RF diode to generate unique IMD signatures, enabling passive RF fingerprinting for device authentication. To emphasize its easy design process and low-cost fabrication, particularly in mass production, the tags were fabricated on both commercially conventional PCB substrate and flexible ultra-thin substrate. Measurement results validated that the increased power at IMD frequencies can be obtained in-band with low transmitting power and the embedded design in lossy material emphasizes the feasibility and suitability for seamless integration into wireless devices.

\bibliographystyle{ieeetr}
\bibliography{areferences}

\begin{thebibliography}{10}

\bibitem{8744656}
C.~Li, L.~Mo, and D.~Zhang, ``Review on uhf rfid localization methods,'' {\em
  IEEE Journal of Radio Frequency Identification}, vol.~3, no.~4, pp.~205--215,
  2019.

\bibitem{4343863}
Y.~Zhao, Y.~Liu, and L.~M. Ni, ``Vire: Active rfid-based localization using
  virtual reference elimination,'' in {\em 2007 International Conference on
  Parallel Processing (ICPP 2007)}, pp.~56--56, 2007.

\bibitem{10008216}
J.~Xu, Z.~Li, K.~Zhang, J.~Yang, N.~Gao, Z.~Zhang, and Z.~Meng, ``The
  principle, methods and recent progress in rfid positioning techniques: A
  review,'' {\em IEEE Journal of Radio Frequency Identification}, vol.~7,
  pp.~50--63, 2023.

\bibitem{8970312}
N.~Soltanieh, Y.~Norouzi, Y.~Yang, and N.~C. Karmakar, ``A review of radio
  frequency fingerprinting techniques,'' {\em IEEE Journal of Radio Frequency
  Identification}, vol.~4, no.~3, pp.~222--233, 2020.

\bibitem{10130767}
J.~He, S.~Huang, Z.~Yang, K.~Yu, H.~Huan, and Z.~Feng, ``Channel-agnostic radio
  frequency fingerprint identification using spectral quotient constellation
  errors,'' {\em IEEE Transactions on Wireless Communications}, vol.~23, no.~1,
  pp.~158--170, 2024.

\bibitem{4211360}
O.~Ureten and N.~Serinken, ``Wireless security through rf fingerprinting,''
  {\em Canadian Journal of Electrical and Computer Engineering}, vol.~32,
  no.~1, pp.~27--33, 2007.

\bibitem{1549967}
K.~Rao, P.~Nikitin, and S.~Lam, ``Antenna design for uhf rfid tags: a review
  and a practical application,'' {\em IEEE Transactions on Antennas and
  Propagation}, vol.~53, no.~12, pp.~3870--3876, 2005.

\bibitem{8736765}
Z.~Khan, X.~Chen, H.~He, J.~Xu, T.~Wang, L.~Cheng, L.~Ukkonen, and J.~Virkki,
  ``Glove-integrated passive uhf rfid tags—fabrication, testing and
  applications,'' {\em IEEE Journal of Radio Frequency Identification}, vol.~3,
  no.~3, pp.~127--132, 2019.

\bibitem{5960759}
F.~Paredes, G.~Zamora, F.~J. Herraiz-Martinez, F.~Martin, and J.~Bonache,
  ``Dual-band uhf-rfid tags based on meander-line antennas loaded with spiral
  resonators,'' {\em IEEE Antennas and Wireless Propagation Letters}, vol.~10,
  pp.~768--771, 2011.

\bibitem{6353124}
A.~A. Babar, T.~Bjorninen, V.~A. Bhagavati, L.~Sydanheimo, P.~Kallio, and
  L.~Ukkonen, ``Small and flexible metal mountable passive uhf rfid tag on
  high-dielectric polymer-ceramic composite substrate,'' {\em IEEE Antennas and
  Wireless Propagation Letters}, vol.~11, pp.~1319--1322, 2012.

\bibitem{7160701}
H.~Bukhari and K.~Sarabandi, ``Miniaturized omnidirectional horizontally
  polarized antenna,'' {\em IEEE Transactions on Antennas and Propagation},
  vol.~63, no.~10, pp.~4280--4285, 2015.

\bibitem{5456169}
H.-W. Liu, C.-F. Yang, and C.-H. Ku, ``Novel miniature monopole tag antenna for
  uhf rfid applications,'' {\em IEEE Antennas and Wireless Propagation
  Letters}, vol.~9, pp.~363--366, 2010.

\bibitem{10414395}
X.~Liu, K.~Ning, S.~Xue, L.~Ge, K.~W. Leung, and J.-F. Mao, ``Printed filtering
  dipole antenna with compact size and high selectivity,'' {\em IEEE
  Transactions on Antennas and Propagation}, vol.~72, no.~3, pp.~2355--2367,
  2024.

\bibitem{5559341}
B.~D. Braaten, ``A novel compact uhf rfid tag antenna designed with series
  connected open complementary split ring resonator (ocsrr) particles,'' {\em
  IEEE Transactions on Antennas and Propagation}, vol.~58, no.~11,
  pp.~3728--3733, 2010.

\bibitem{99054}
J.~Rashed and C.-T. Tai, ``A new class of resonant antennas,'' {\em IEEE
  Transactions on Antennas and Propagation}, vol.~39, no.~9, pp.~1428--1430,
  1991.

\bibitem{6171817}
O.~O. Olaode, W.~D. Palmer, and W.~T. Joines, ``Characterization of meander
  dipole antennas with a geometry-based, frequency-independent lumped element
  model,'' {\em IEEE Antennas and Wireless Propagation Letters}, vol.~11,
  pp.~346--349, 2012.

\bibitem{1504825}
C.-C. Lin, S.-W. Kuo, and H.-R. Chuang, ``A 2.4-ghz printed meander-line
  antenna for usb wlan with notebook-pc housing,'' {\em IEEE Microwave and
  Wireless Components Letters}, vol.~15, no.~9, pp.~546--548, 2005.

\bibitem{9263312}
A.~S.~M. Alqadami, A.~E. Stancombe, K.~S. Bialkowski, and A.~Abbosh, ``Flexible
  meander-line antenna array for wearable electromagnetic head imaging,'' {\em
  IEEE Transactions on Antennas and Propagation}, vol.~69, no.~7,
  pp.~4206--4211, 2021.

\bibitem{9128053}
T.~T. Le and T.-Y. Yun, ``Miniaturization of a dual-band wearable antenna for
  wban applications,'' {\em IEEE Antennas and Wireless Propagation Letters},
  vol.~19, no.~8, pp.~1452--1456, 2020.

\bibitem{6335459}
Z.~L. Ma, L.~J. Jiang, J.~Xi, and T.~T. Ye, ``A single-layer compact hf-uhf
  dual-band rfid tag antenna,'' {\em IEEE Antennas and Wireless Propagation
  Letters}, vol.~11, pp.~1257--1260, 2012.

\bibitem{5299014}
B.~D. Braaten, M.~Reich, and J.~Glower, ``A compact meander-line uhf rfid tag
  antenna loaded with elements found in right/left-handed coplanar waveguide
  structures,'' {\em IEEE Antennas and Wireless Propagation Letters}, vol.~8,
  pp.~1158--1161, 2009.

\bibitem{6648416}
G.~A. Casula, G.~Montisci, and G.~Mazzarella, ``A wideband pet inkjet-printed
  antenna for uhf rfid,'' {\em IEEE Antennas and Wireless Propagation Letters},
  vol.~12, pp.~1400--1403, 2013.

\bibitem{9591355}
P.~Wang, W.~Luo, Y.~Shao, and H.~Jin, ``An uhf rfid circularly polarized tag
  antenna with long read distance for metal objects,'' {\em IEEE Antennas and
  Wireless Propagation Letters}, vol.~21, no.~2, pp.~217--221, 2022.

\bibitem{7748548}
Y.~Yao, C.~Cui, J.~Yu, and X.~Chen, ``A meander line uhf rfid reader antenna
  for near-field applications,'' {\em IEEE Transactions on Antennas and
  Propagation}, vol.~65, no.~1, pp.~82--91, 2017.

\bibitem{7384427}
T.~Leng, X.~Huang, K.~Chang, J.~Chen, M.~A. Abdalla, and Z.~Hu, ``Graphene
  nanoflakes printed flexible meandered-line dipole antenna on paper substrate
  for low-cost rfid and sensing applications,'' {\em IEEE Antennas and Wireless
  Propagation Letters}, vol.~15, pp.~1565--1568, 2016.

\bibitem{5979187}
T.~Zhang, R.~Li, G.~Jin, G.~Wei, and M.~M. Tentzeris, ``A novel multiband
  planar antenna for gsm/umts/lte/zigbee/rfid mobile devices,'' {\em IEEE
  Transactions on Antennas and Propagation}, vol.~59, no.~11, pp.~4209--4214,
  2011.

\bibitem{6843861}
T.~Pan, S.~Zhang, and S.~He, ``Compact rfid tag antenna with circular
  polarization and embedded feed network for metallic objects,'' {\em IEEE
  Antennas and Wireless Propagation Letters}, vol.~13, pp.~1271--1274, 2014.

\bibitem{8758307}
H.~M. Santos, P.~Pinho, R.~P. Silva, M.~Pinheiro, and H.~M. Salgado,
  ``Meander-line monopole antenna with compact ground plane for a bluetooth
  system-in-package,'' {\em IEEE Antennas and Wireless Propagation Letters},
  vol.~18, no.~11, pp.~2379--2383, 2019.

\bibitem{4020418}
C.~T. Rodenbeck, ``Planar miniature rfid antennas suitable for integration with
  batteries,'' {\em IEEE Transactions on Antennas and Propagation}, vol.~54,
  no.~12, pp.~3700--3706, 2006.

\bibitem{pozar2021microwave}
D.~M. Pozar, {\em Microwave engineering: theory and techniques}.
\newblock John wiley \& sons, 2021.

\bibitem{lin2011introduction}
M.-B. Lin, {\em Introduction to VLSI systems: a logic, circuit, and system
  perspective}.
\newblock CRC press, 2011.

\end{thebibliography}

\end{document}